\documentclass[aps,pra,twocolumn,showpacs,superscriptaddress,amsmath,amssymb,floatfix]{revtex4-1}
\usepackage{epsfig}
\usepackage{graphicx}
\usepackage{bm}
\usepackage{subfigure}

\begin{document}
\title{Three-terminal heat engine and refrigerator based on superlattices}

\author{Yunjin Choi}
\affiliation{Department of Physics and Astronomy \& Rochester Theory Center, University of Rochester, Rochester, New York 14627, USA}

\author{Andrew N. Jordan}
\email{jordan@pas.rochester.edu}
\affiliation{Department of Physics and Astronomy \& Rochester Theory Center, University of Rochester, Rochester, New York 14627, USA}

\affiliation{Institute for Quantum Studies, Chapman University, 1 University Drive, Orange, CA 92866, USA}

\date{\today}
\begin{abstract}
We propose a three terminal heat engine based on semiconductor superlattices for energy harvesting. The periodicity of the superlattice structure creates an energy miniband, giving an energy window for allowed electron transport. We find that this device delivers a large power, nearly twice than the heat engine based on quantum wells, with a small reduction of efficiency. This engine also works as a refrigerator in a different regime of the system's parameters. The thermoelectric performance of the refrigerator is analyzed, including the cooling power and coefficient of performance in the optimized condition. We also calculate phonon heat current through the system, and explore the reduction of phonon heat current compared to the bulk material. The direct phonon heat current is negligible at low temperatures, but dominates over the electronic at room temperature and we discuss ways to reduce it.
\end{abstract}

\maketitle

\section{Introduction}
There has been increasing interest in developing high efficiency, high power thermoelectric devices, constructed from the bottom-up using nanoscale designs. The primary applications driving interest in this area are energy-harvesting, the collection and conversion of waste heat to electrical power, produced from sources ranging from hand-held electronics to industrial sources of heat, and refrigeration, actively cooling a spatial region via electrons to evacuate heat out of an area. The use of nanoscale architecture instead of bulk materials is motivated by the low figure of merit - or poor thermoelectric conversion efficiency - of bulk materials, whereas conduction through nanoscale electronics can reach Carnot efficiency.

One way to produce high thermodynamic efficiency in the conversion of heat to power is the use of structures with sharp spectral features, such as quantum dots \cite{Hicks,Hicks2,Mahan}. The use of quantum dots in thermoelectric transport has been extensively researched in the past several years \cite{Beenakker,Humphrey,Kubala,Billings,Nakpathomkun,Jordan,Kennes,Weymann}.  See Ref.~\cite{Sothmann,Benenti} for recent reviews of these and related activities. In particular, if the dot is transporting electrons via resonant tunneling, the quantum dot acts as an energy filter - permitting the ``tight-coupling" of heat and charge transport which can lead to Carnot efficiency. Other structures from mesoscopic physics, including the quantum point contact and electron cavity \cite{Sothmann1}, quantum wells \cite{Sothmann2,Agarwal}, quantum Hall bar \cite{Sothmann4,Sanchez2}, superconducting leads \cite{Mazza2}, and Coulomb blockaded quantum dots(s) \cite{Sanchez3,Sanchez,Jiang,Mazza1,Henriet} have also been investigated for their multi-terminal thermoelectric properties.  The late Prof. Markus B\"uttiker, for whom this special issue is in memory of, was highly influential in the theoretical development of these ideas, as can be seen in the above list of references.

Several experiments have begun exploring this physics. Prance \textit{et al}. \cite{Prance} performed experiments on a cavity connected to resonant tunneling quantum dots acting as an electronic refrigerator, based on the proposal of Edwards \textit{et al.} \cite{Edwards}. They demonstrated that applying bias to the system results in cooling a large $6\mu m^2$ cavity from $280mK$ to below $190 mK$. Very recently, Roche \textit{et al.} \cite{Roche} and F. Hartmann \textit{et al.} \cite{Hartmann} showed rectification of electrical current of the nano Amp scale and power production on a pico Watt scale from a capacitively coupled source of fluctuations. This was based on the theoretical proposal of Sothmann \textit{et al.} \cite{Sothmann1}.

While a nanoscale thermoelectric generator can power nanoscale devices, it is of great interest to find practical ways to scale up these nanoengines. One way is to simply add them in electrical series while being able to couple to a common source of heat. In commercial thermoelectric generators, this is usually done by alternating the semiconductor type, of either p-type or n-type to be able to apply the heat difference in parallel because the heat and electrical transport are in opposite directions in a p-type semiconductor \cite{DiSalvo}.  This permits the generated voltage to grow with the number of elements, while keeping the current fixed.  Various other ways of scaling the devices have been proposed \cite{Harman,Chena,Yu,Bell}. In Jordan \textit{et al.}, a layered structure was proposed by alternating layers of semiconductor and self-assembled quantum dots, so as to create a large-scale device where heat and electrical transport are separated, while keeping the high thermodynamic efficiency \cite{Jordan}.  This is a parallel strategy of scaling, so the generated current grows with the number of dots, while the voltage difference is fixed.
Sothmann \textit{et al.} considered a technically simpler method of creating quantum wells that permit resonant tunneling \cite{Sothmann2}. The physics there is somewhat different because energy may be distributed into the transverse degrees of the electron motion. Nevertheless, reasonable thermodynamic efficiency was found, with increased power production.

One of the outstanding challenges to creating high-efficiency thermoelectric devices is phonon transport. Phonons give a way for the hot and cold side of the device to exchange energy directly, without converting it to power via the electrons. Therefore, any possible way to reduce the phonon transport while still allowing electron transport will aid in the overall thermodynamic efficiency. Interface-based devices, such as described above can help with this, because the interface helps to reflect the phonons \cite{Colvard,Tamura1,Hyldgaard,Chen,Tamura2,Simkin,Venkat1,Venkat2}. Ideally, there will be additional material layers that act as thermal insulators.

The purpose of the present article is to build on these accomplishments, and make an analysis of a thermoelectric device based on semiconductor super-lattices. These structures are fabricated by making a periodic layered structure of alternating materials, such as GaAs/AlGaAs. The effect on the electronic transport is to form a series of mini-bands of allowed and forbidden energies where conduction electrons can transport \cite{Tsu,Esaki,Smith,Wacker}. This structure can be considered as a generalization of the resonant tunneling quantum wells. The mini-band gives a top-hat profile of variable width for the energy-filtering.
Such a top-hat profile has been argued by Whitney to offer the highest efficiency for a given power extraction \cite{Whitney2014}.  However, we note the transverse degrees of freedom make the system somewhat different.
At a small band width, our system will be similar to a quantum well, but can be extended to allow a fixed width longitudinal energy window. We make a first-principles analysis of the heat and charge transport in a three-terminal geometry, where two terminals carry charge, and a third carries heat (see Fig.~\ref{fig_gemoetry_energydiagram}). The offset of the miniband centers and their respective widths determine the power produced and efficiency of the heat conversion given fixed temperature differences. We next make an analysis of the coefficient of performance of this device for the purposes of refrigeration of the central region. The final purpose of the present work  is to also make a systematic calculation of the heat transport due to phonons.  We make a detailed investigate the heat current through the device from phonon transport using a Kronig-Penney model, and consider different ways to stop it.

The paper is organized as follows. In section \ref{heat_engine_electron}, we introduce a model of the superlattice heat engine in its dual roles: the energy harvester in section \ref{energy_harvesting} and the refrigerator in section \ref{refrigeration}.  We discuss our results for the generated power and the efficiency of the engine in section \ref{result}, and show the cooling power and the coefficient of performance in section \ref{refrigeration}. The second part, section \ref{heat_engine_phonon} focuses on phonon heat current generated by the heat engine and discusses the effect on the efficiency of the heat engine. We finish with our conclusions in section \ref{conclusion}.

\section{Heat engine based on superlattices}\label{heat_engine_electron}
\subsection{Energy harvesting by electron transport}\label{energy_harvesting}
\begin{figure}
\begin{center}
\subfigure{\includegraphics[width=8.5cm]{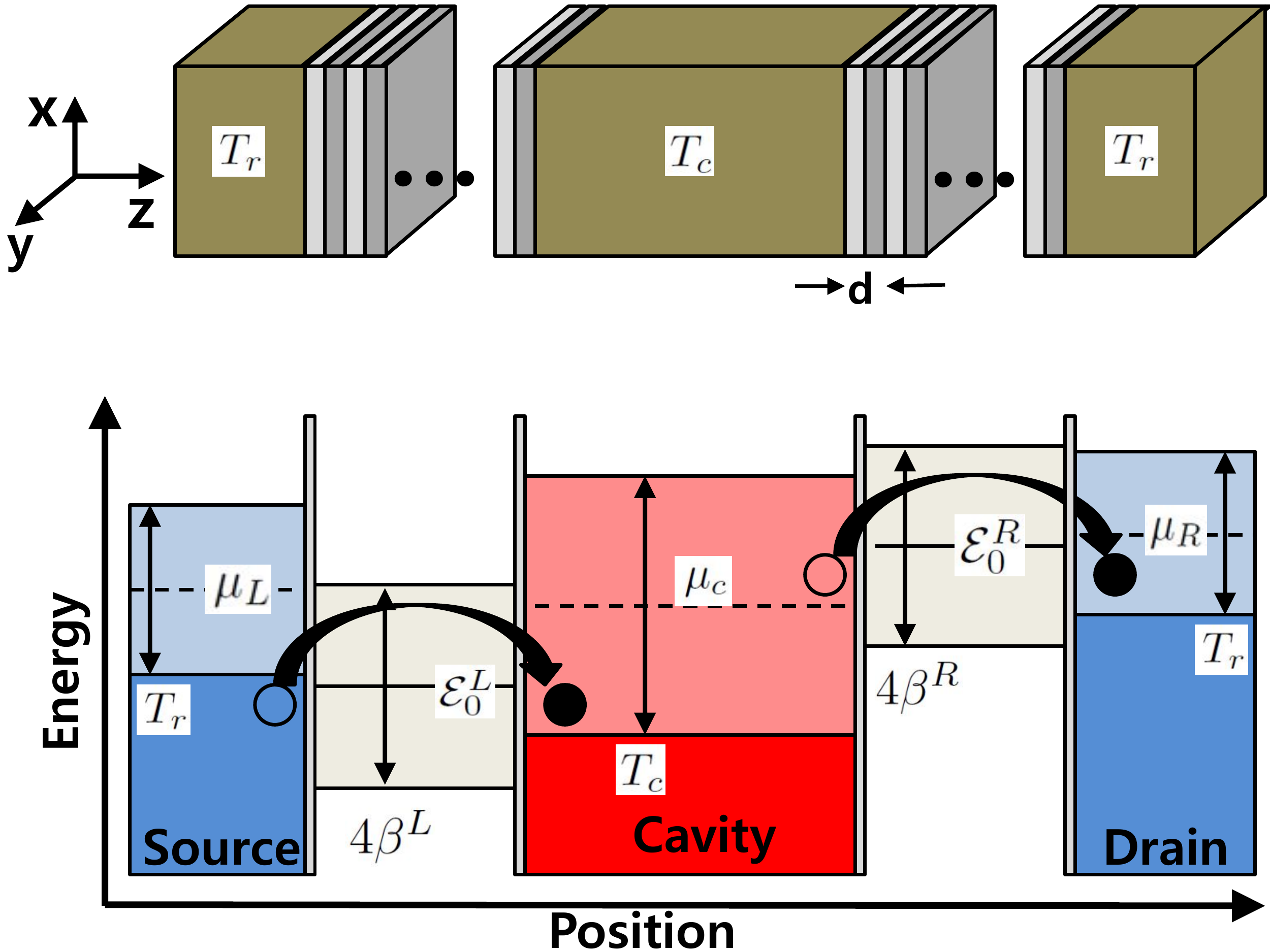}}
\caption{(Top) Schematic of the superlattice heat engine. A hot cavity at temperature $T_{c}$ is coupled via superlattices to cold electronic reservoirs at temperature $T_r$. (Bottom) The periodic structure of the superlattices form the miniband centered at $\mathcal{E}_0^{L/R}$ with the width $4\beta^{L/R}$ when we apply the bias voltage $\mu_R-\mu_L=eV$. The gray shading area shows the energy miniband where the electrons can transport between regions. The shadings in the source, cavity, and drain regions indicate thermal smearing. }\label{fig_gemoetry_energydiagram}
\end{center}
\end{figure}

We consider a setup shown schematically in Fig.~\ref{fig_gemoetry_energydiagram}. It consists of a center cavity connected to two electronic reservoirs $r=L,R$ via a superlattice.
The electronic reservoirs are characterized by the occupation of the states given by the Fermi function, $f_r(E)=[\exp[(E-\mu_r)/(k_BT_r)]+1]^{-1}$ with temperature $T_r$ and chemical potential $\mu_r$, and the underlying assumption is that inelastic scattering processes restore the local thermal equilibrium on a fast time scale. The center cavity is also assumed to be in thermal equilibrium with a heat bath of temperature $T_{c}$.
We assume the structure is translationally symmetric within the $x$ and $y$ directions, perpendicular to the growth direction $z$.
The superlattices are designed as periodic structures with lattice constant $d$ (sum of the width of well and barrier thickness). The periodicity of the structure in the $z$ direction implies that the eigenstates of the Hamiltonian can be written as Bloch states with the Bloch vector $q\in[-\pi/d,\pi/d]$, so the simple solutions with the Kronig-Penney model resemble the standard superlattices \cite{Esaki}. The corresponding eigenvalues of the Hamiltonian form a miniband. The allowed energies of the miniband can be written,
\begin{eqnarray}
E^{\nu}(q)=\mathcal{E}_0^{\nu}-2\beta^{\nu}\cos qd,\label{superlattice_energy}
\end{eqnarray}
where this is the result of a standard tight-binding calculation with band indices $\nu$. $\mathcal{E}_0^{\nu}$ is a center of miniband $\nu$ and its width is $4\beta^{\nu}$. Our discussion is restricted to the electron transport through the lowest miniband, thus the miniband index $\nu$ is neglected from here on.

To find the electric and heat currents through the superlattices expressed in terms of an integral over energy, we first find the density of states of the superlattice. For the given miniband, the energy is that of a two dimensional electron gas with the bottom of the band at $E(q)$ in Eq.~(\ref{superlattice_energy}). Therefore the three dimensional density of states is given by \cite{Davies}
\begin{eqnarray}
\nu_{3D}(E)&=&\frac{1}{2}\int_{-\infty}^{\infty}d\varepsilon\;\nu_{SL}(\varepsilon)\;\nu_{2D}(E-\varepsilon)\nonumber\\
&=&\frac{\nu_{2D}}{2}\int_{-\infty}^{E}d\varepsilon\;\nu_{SL}(\varepsilon).
\end{eqnarray}
The factor $1/2$ is to avoid double counting the spin and $\nu_{2D}=m/\pi\hbar^2$ is a two dimensional density of states per unit area. The density of states of the superlattice can be factorized into longitudinal and transverse  parts, and the one dimensional superlattice density of states $\nu_{SL}$ is
\begin{eqnarray}
\nu_{SL}(E)=\frac{1}{\pi\beta d}\frac{\Theta(E-E_z^-)\Theta(-E+E_z^+)}{\sqrt{1-(\frac{-E+\mathcal{E}_0}{2 \beta})^2}},\label{density_of_state}
\end{eqnarray}
where we use a Heaviside step function $\Theta$ to show the range of the energy $E$ with the maximum/miminum value $E_z^{\pm}=\mathcal{E}_0\pm 2\beta$. Therefore, the electrons only in selected values of energy $E_z^-<E<E_z^+$ will transport and generate current. We assume the central cavity region is strongly coupled to the external source of energy, so the occupation is described as a Fermi function with local temperature $T_c$ determined by the thermal reservoir.

The electric and energy currents for simplified miniband transport in $z$ direction emitted by the reservoir $r$ into an cavity $c$
can be evaluated within a Landauer-B\"{u}ttiker approach as \begin{eqnarray}
I_r&=&\frac{e\nu_{2D}\mathcal{A}}{2\pi\hbar}\int_{E_{z}^{r-}}^{E_{z}^{r+}}dE_{z}dE_{\bot}[f_r(E)-f_c(E)], \label{current_r}\\
J_r&=&\frac{\nu_{2D}\mathcal{A}}{2\pi\hbar}\int_{E_{z}^{r-}}^{E_{z}^{r+}}dE_{z}dE_{\perp}\;E\;[f_r(E)-f_c(E)],\label{heatcurrent_r}
\end{eqnarray}
where $\mathcal{A}$ is the surface area of the superlattice, $E_{z}^{r\pm}$ is the maximum/minimum energy of the reservoir $r$, and we denote $E_\perp$ as the energy carried in the transverse degrees of freedom, and $E_z$ as the energy carried in the longitudinal degree of freedom, so that $E = E_\perp+E_z$. Here, the square root in Eq. (\ref{density_of_state}) cancels the velocity of the electron, $v_g$, in the current $I_r=(e/4)\int dE_z v_g\nu_{3D}[f_r(E)-f_{c}(E)]$ \cite{exp}.
We see that the range of the integral comes from the density of states which gives a transmission function of flat box form,
$\mathcal{T}^{r}(E_z)=\Theta(E_z-E_z^{r-})\Theta(-E_z+E_z^{r+})$. To rewrite the above equations, we introduce the integrals $K_1(x)=\int_0^{\infty}dt (1+e^{t-x})^{-1}=\log(1+e^x)$ and
$K_2(x)=\int_0^{\infty}dt\: t(1+e^{t-x})^{-1}=-\textmd{Li}_2(-e^x)$ with
the dilogarithm $\textmd{Li}_2(z)=\sum_{k=1}^{\infty}\frac{z^k}{k^2}$. Then the simplified analytic expressions of the electric and energy currents are
\begin{eqnarray}
I_r&=&\frac{e\nu_{2D}\mathcal{A}k_B}{2\pi\hbar}\int_{E_{z}^{r-}}^{E_{z}^{r+}}dE_{z}
\left[T_rK_1(\tilde{E_z^r})-T_cK_1(\tilde{E_z^c}) \right], \label{electric_current_general}\\
J_r&=&\frac{\nu_{2D}\mathcal{A}k_B}{2\pi\hbar} \left[\int_{E_{z}^{r-}}^{E_{z}^{r+}}dE_{z}E_z
\left[  T_rK_1(\tilde{E_z^r})-T_cK_1(\tilde{E_z^c})  \right]\right. \nonumber\\
&&\left.+k_B^2\int_{E_{z}^{r-}}^{E_{z}^{r+}}dE_{z} \left[  T_r^2K_2(\tilde{E_z^r})
-T_c^2K_2(\tilde{E_z^c})\right] \right],\label{heat_current_general}
\end{eqnarray}
where $\tilde{E_z^r}=(\mu_r-E_z)/k_BT_r$ and $\tilde{E_z^c}=(\mu_c-E_z)/k_BT_c$.

\subsection{Results}\label{result}
We now analyze the system by focusing on linear response and later turn to the nonlinear regime. To simplify the analysis of the system, we introduce the average temperature $T=(T_r+T_c)/2$ and the temperature difference
$\Delta T=T_c-T_r$. For energy harvesting, the temperature difference is considered to be $\Delta T>0$. (If we consider refrigeration, the temperature difference is $\Delta T<0$.) We introduce the bias $\mu_R-\mu_L=eV$ to the system by applying $\mu_R=eV/2$ and $\mu_L=-eV/2$.
We also rewrite the width of each miniband, $\beta^L=\beta+\alpha$ and $\beta^R=\beta-\alpha$ where $\alpha$, with $|\alpha|<\beta$, determines the asymmetry between the left and right energy width, so the relative thickness of the left and right miniband width is determined by $\alpha$.

The chemical potential of the cavity $\mu_c$ as well as the temperature $T_c$ are determined by imposing conservation of charge and energy, $I_L+I_R=0$ and  $J+J_L+J_R=0$ where $J$ is the heat current entering from the heat source. From these conservation laws, we can obtain the electric and heat currents through the system, as well as $J$ and $\mu_c$.

\subsubsection{Linear response}
To linear order in the temperature difference $\Delta T$ and the bias voltage $V$, the net current flowing through the system, $I_L=-I_R\equiv I$, is given by
\begin{eqnarray}
I&=& G_e V+G_e S_t\Delta T.\label{current_linear}
\end{eqnarray}
The electrical conductance $G_e$ and thermopower (or Seebeck coefficient) $S_t$ of the system are
\begin{eqnarray}
G_e&=&-\frac{e^2\nu_{2D}\mathcal{A}}{2\pi\hbar}\frac{G_{L1}G_{R1}}{G_{L1}+G_{R1}},\label{electric_conductance}\\
S_t&=&\frac{k_B}{e}\left[\frac{G_{L2}+G_{L3}}{G_{L1}}- \frac{G_{R2}+G_{R3}}{G_{R1}} \right],\label{thermopower}
\end{eqnarray}
where we have introduced the auxiliary functions
\begin{eqnarray}
G_{r1}&=&\int_{E_z^{r-}}^{E_z^{r+}}dE_z\frac{1}{1+e^{E_z/k_BT}},\label{gr1}\\
G_{r2}&=&\int_{E_z^{r-}}^{E_z^{r+}}dE_z\frac{E_z/k_BT}{1+e^{E_z/k_BT}},\label{gr2}\\
G_{r3}&=&\int_{E_z^{r-}}^{E_z^{r+}}dE_z\log(1+e^{-E_z/k_BT}).\label{gr3}
\end{eqnarray}
The electrical conductance shows that $G_{L1}$($G_{R1}$) is proportional to the electrical conductance of the left(right) superlattice, so the net conductance $G_e$ is simply the series combination of the two conductors. The three-terminal thermopower $S_t$ is determined by the difference between left and right two-terminal thermopower of each superlattice, and vanishes when they are identical \cite{Sothmann2,Sanchez2}. Therefore, depending on the magnitude of the left and right properties, $S_t$ shows whether the system is analogous to a p-type or n-type semiconductor. When $S_t$ is positive for $(G_{L2}+G_{L3})/G_{L1}> (G_{R2}+G_{R3})/G_{R1}$, the system acts as if the mobile charge carrier is positive and behaves like a p-type semiconductor, and vice versa.

\begin{figure}
\begin{center}
\subfigure{\includegraphics[width=4.2cm]{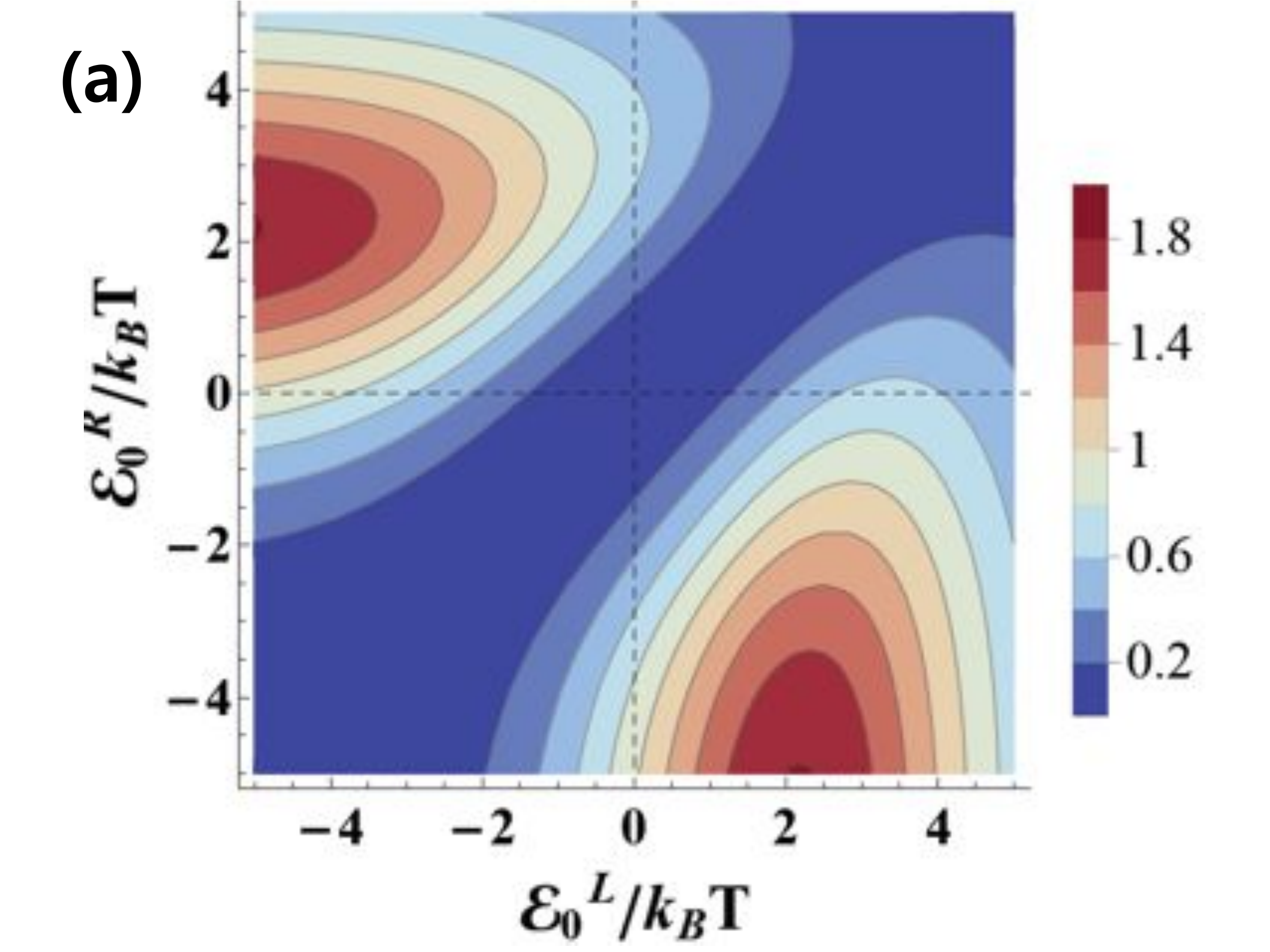}}
\subfigure{\includegraphics[width=4.2cm]{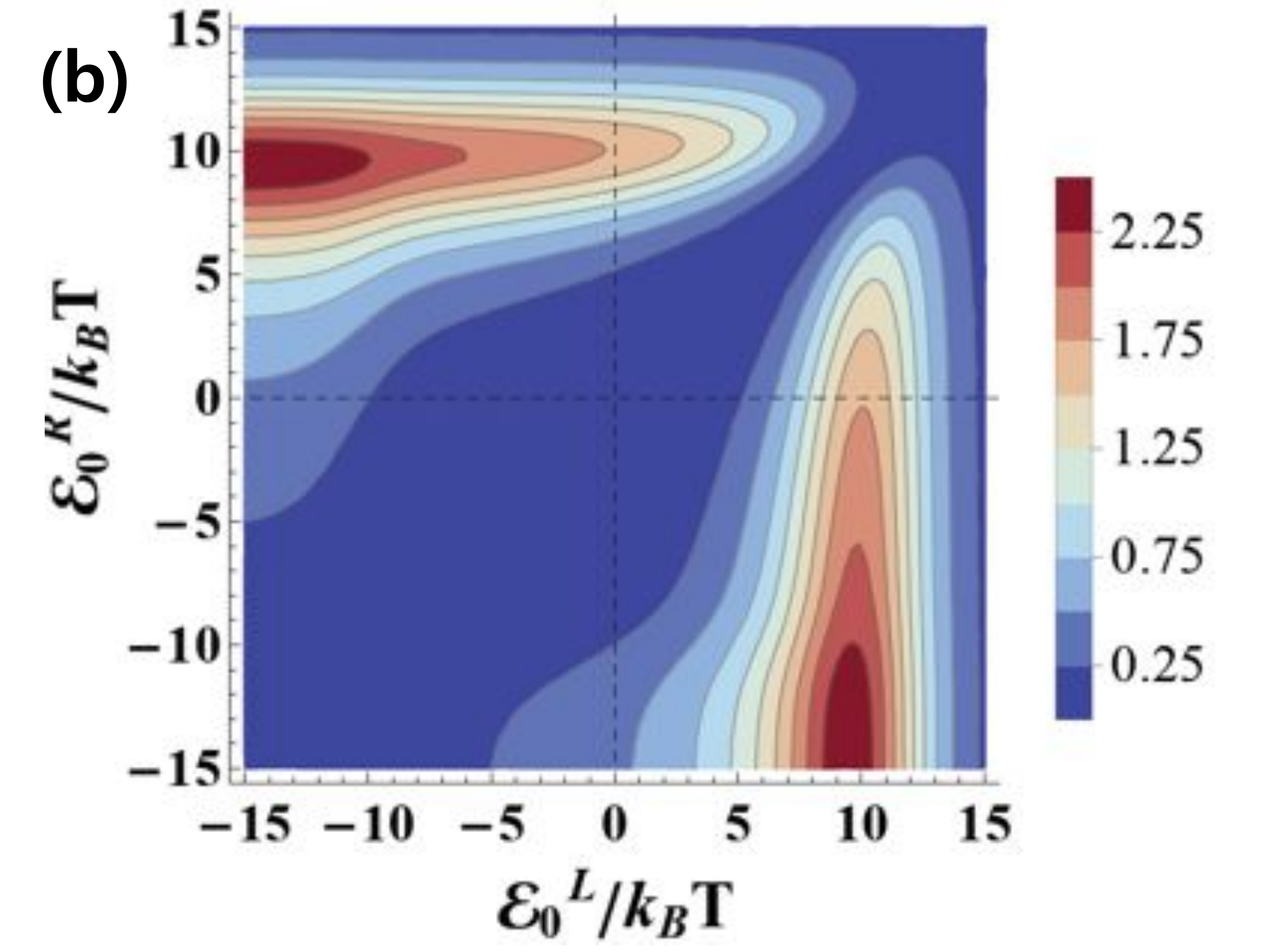}}
\subfigure{\includegraphics[width=4.2cm]{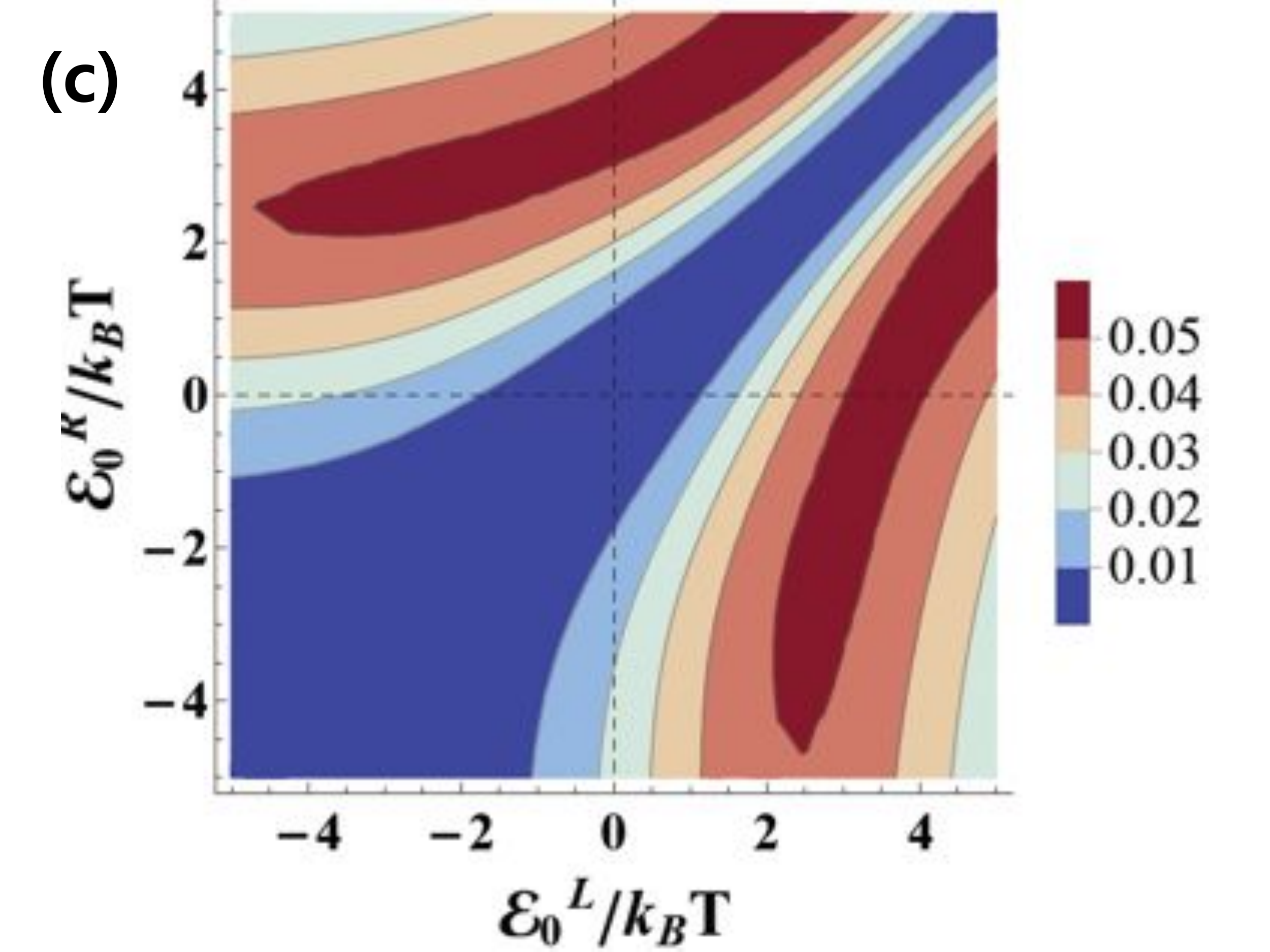}}
\subfigure{\includegraphics[width=4.2cm]{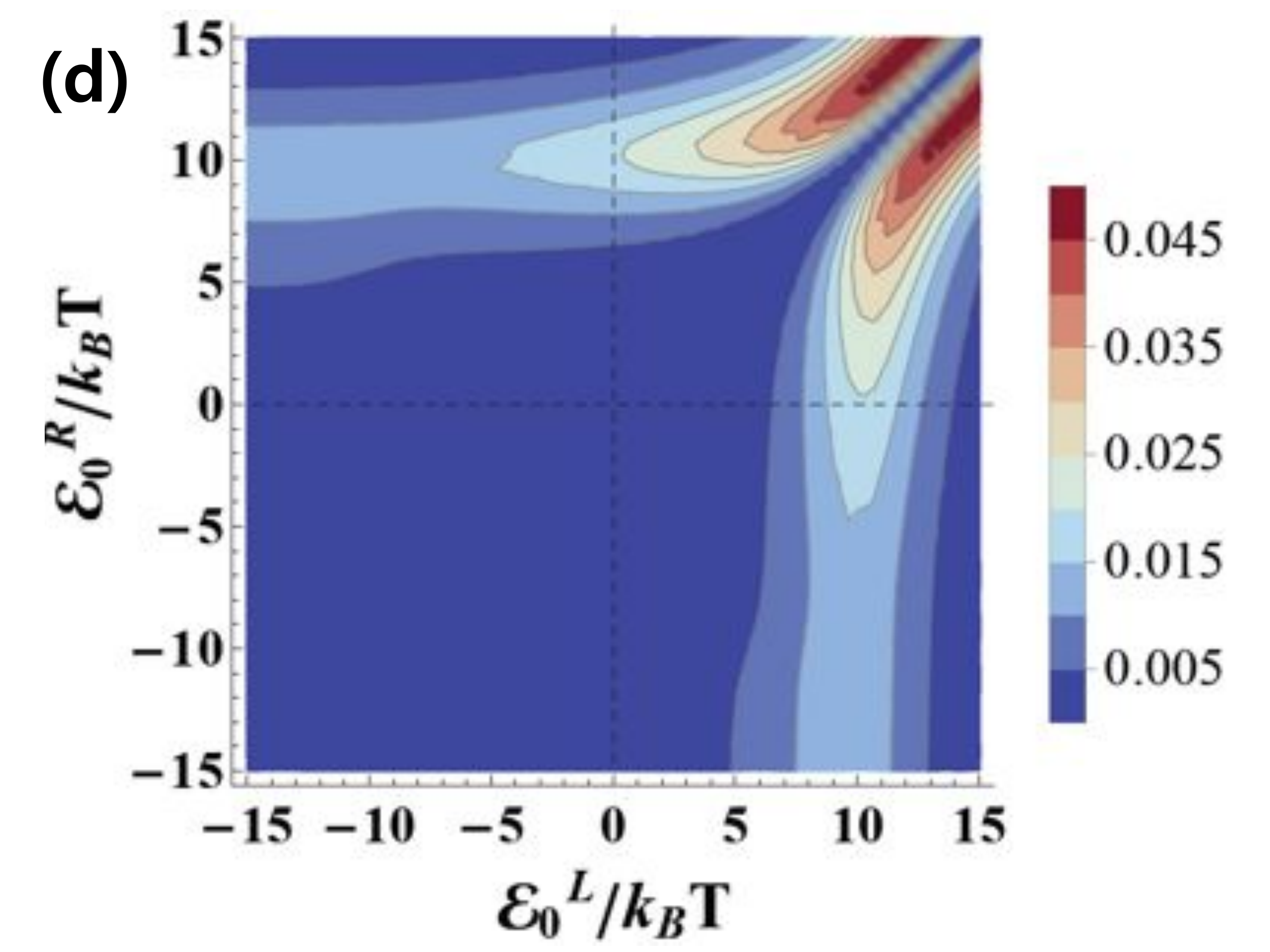}}
\caption{(a), (b) Maximum power in units of $\frac{\nu_{2D}\mathcal{A}(k_BT)}{2\hbar}(\frac{k_B\Delta T}{2})^2$ within linear response with respect to the centers of each miniband for a symmetric configuration, $\alpha=0$, when (a) is $\beta=k_BT$ and (b) is $\beta=5k_BT$. (c), (d) Efficiency at maximum power normalized by Carnot efficiency $\eta_c$ within linear response as a function of the centers of each miniband for symmetric setup. Panel (c)/(d) is a corresponding efficiency for the case of (a)/(b). All plots are obtained for $T=300K$ and $\Delta T=1K$.}\label{fig_linear_maxpower}
\end{center}
\end{figure}
The bias voltage $V$ applied against heat driven charge current generates a finite output power $P=|IV|$. The power vanishes either when no bias voltage is applied or at the stopping voltage, $V_{stop}=-S_t\Delta T$. The output power takes its maximum value at half of the stopping voltage,
\begin{eqnarray}
P_{max}=|G_e|\frac{(S_t\Delta T)^2}{4}.\label{maxpower_linear}
\end{eqnarray}
In order to evaluate the efficiency $\eta$ given by the ratio of output power to input heat current, we need to find the heat current $J=-J_L-J_R$ injected from the heat bath,
\begin{eqnarray}
J=G_e\Pi\: V+(G_eS_t\Pi+H_{t1}+H_{t2})\Delta T\label{heatcurrent_linear},
\end{eqnarray}
which shows the Peltier effect.  Here the coefficients $\Pi$, $H_{t1}$, and $H_{t2}$ are defined as
\begin{eqnarray}
\Pi&=&\frac{k_BT}{e}\left[ \frac{G_{R2}+G_{R3}}{G_{R1}}-\frac{G_{L2}+G_{L3}}{G_{L1}}  \right],\label{Peltier_coeff}\\
H_{t1}&=&\frac{\nu_{2D}\mathcal{A}}{2\pi\hbar}\:k_B^2T\frac{G_{L1}-G_{R1}}{G_{L1}+G_{R1}} \left[\frac{(G_{R2}+G_{R3})^2}{G_{R1}} \right.\nonumber\\ &&\left.-\frac{(G_{L2}+G_{L3})^2}{G_{L1}}-4(G_{R2}+G_{R3})(G_{L2}+G_{L3}) \right],\nonumber\\ \label{H_t1}\\
H_{t2}&=&\frac{\nu_{2D}\mathcal{A}}{2\pi\hbar}\:k_B^2T \left[ (G_{L4}+2G_{L5}-G_{L6})\right.\nonumber\\
&&\left.+(G_{R4}+2G_{R5}-G_{R6}) \right],\label{H_t2}
\end{eqnarray}
together with the auxiliary functions
\begin{eqnarray}
G_{r4}&=&\int_{E_z^{r-}}^{E_z^{r+}}dE_z\frac{(E_z/k_BT)^2}{1+e^{E_z/k_BT}},\\
G_{r5}&=&\int_{E_z^{r-}}^{E_z^{r+}}dE_z\frac{E_z}{k_BT}\log(1+e^{-E_z/k_BT}), \\
G_{r6}&=&\int_{E_z^{r-}}^{E_z^{r+}}dE_z\textmd{Li}_2(-e^{-E_z/k_BT}).
\end{eqnarray}

Rewriting the heat current in Eq. (\ref{heatcurrent_linear}) as $J=\Pi I+(H_{t1}+H_{t2})\Delta T$, we see the meaning of the coefficients more clearly. The first part is simply proportional to the charge current, and $\Pi$ shows the presence of heating from the Peltier contribution. The relation $\Pi=-TS_{t}$ from Eq.~(\ref{thermopower}) and Eq.~(\ref{Peltier_coeff}) shows the coefficient $\Pi$ can be considered as the back action counterpart to $S_t$, and the minus sign of the relation comes because $J$ is the heat current injected from the heat bath into the cavity. This relation shows our system satisfies Onsager symmetry resulting from the time reversibility, which relates the Seeback and Peltier coefficients \cite{Onsager1931}. The second part shows the heat generated by the temperature difference and the thermal conductance is $H_{t1}+H_{t2}$. The first one, $H_{t1}$, contributes to the heat flow from the asymmetric superlattices and disappears when the left and right superlattices are symmetric.

The heat current at maximum power takes the form
\begin{eqnarray}
J_{maxP}=\left[ \frac{G_eS_t\Pi}{2}+H_{t1}+H_{t2} \right]\Delta T.
\end{eqnarray}
Therefore, the efficiency at maximum power is given by
\begin{eqnarray}
\eta_{maxP}&=&\left|\frac{G_eS_t^2}{2G_eS_t\Pi+4H_{t1}+4H_{t2}}\right|\Delta T\nonumber\\
&\approx&\left| \frac{G_eS_t^2 T}{2G_eS_t\Pi+4H_{t1}+4H_{t2}}\right|\:\eta_c,\label{efficiency_maxP}
\end{eqnarray}
where the second approximation comes from the Carnot efficiency for the small temperature difference, $\eta_c=\Delta T/T_c \approx\Delta T/T$.  Therefore, the combination of the coefficients, inside of the vertical bars in Eq.~(\ref{efficiency_maxP}) determines the efficiency ratio to Carnot efficiency.

\begin{figure}
\begin{center}
\subfigure{\includegraphics[width=4.2cm]{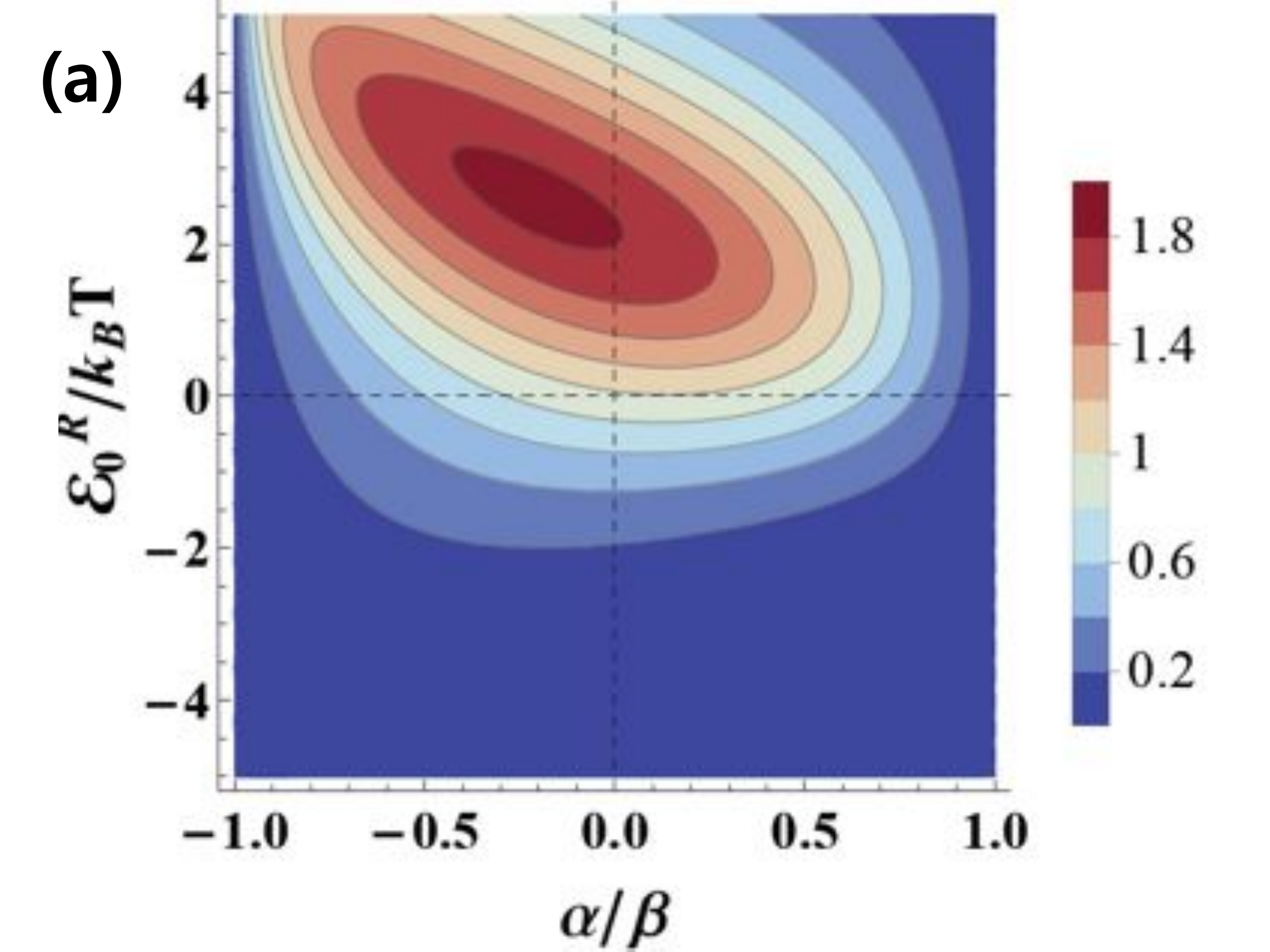}}
\subfigure{\includegraphics[width=4.2cm]{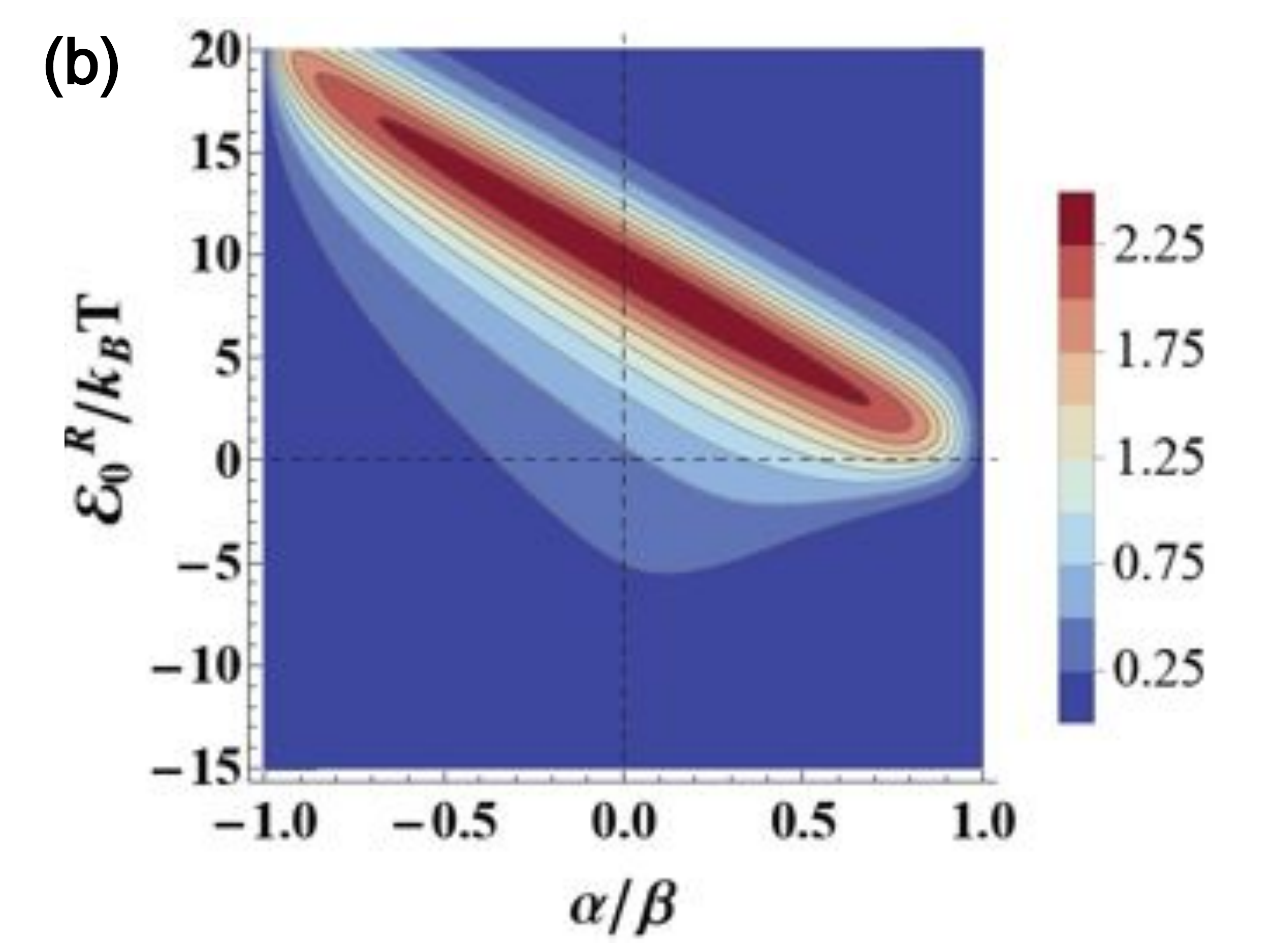}}
\subfigure{\includegraphics[width=4.2cm]{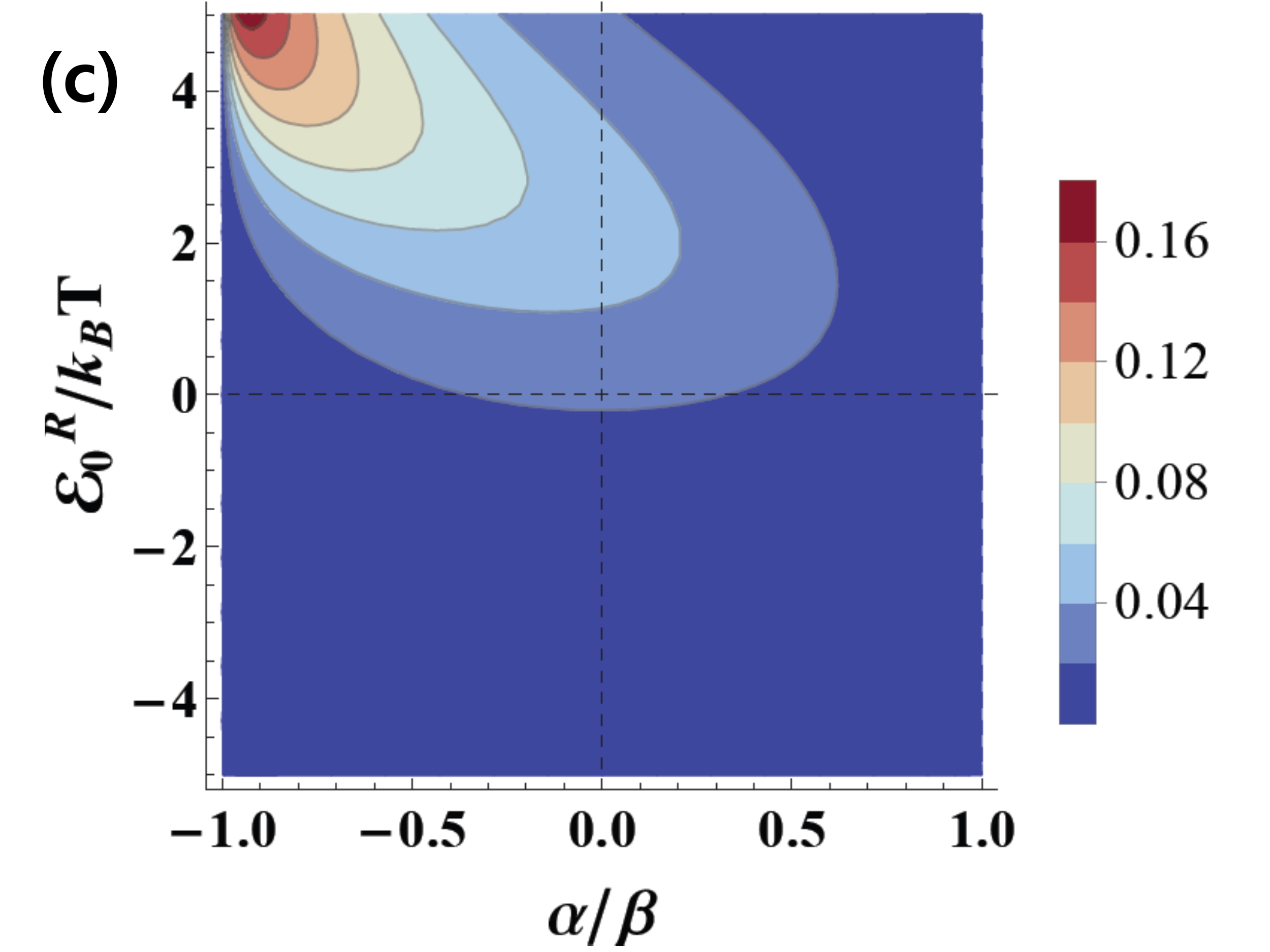}}
\subfigure{\includegraphics[width=4.2cm]{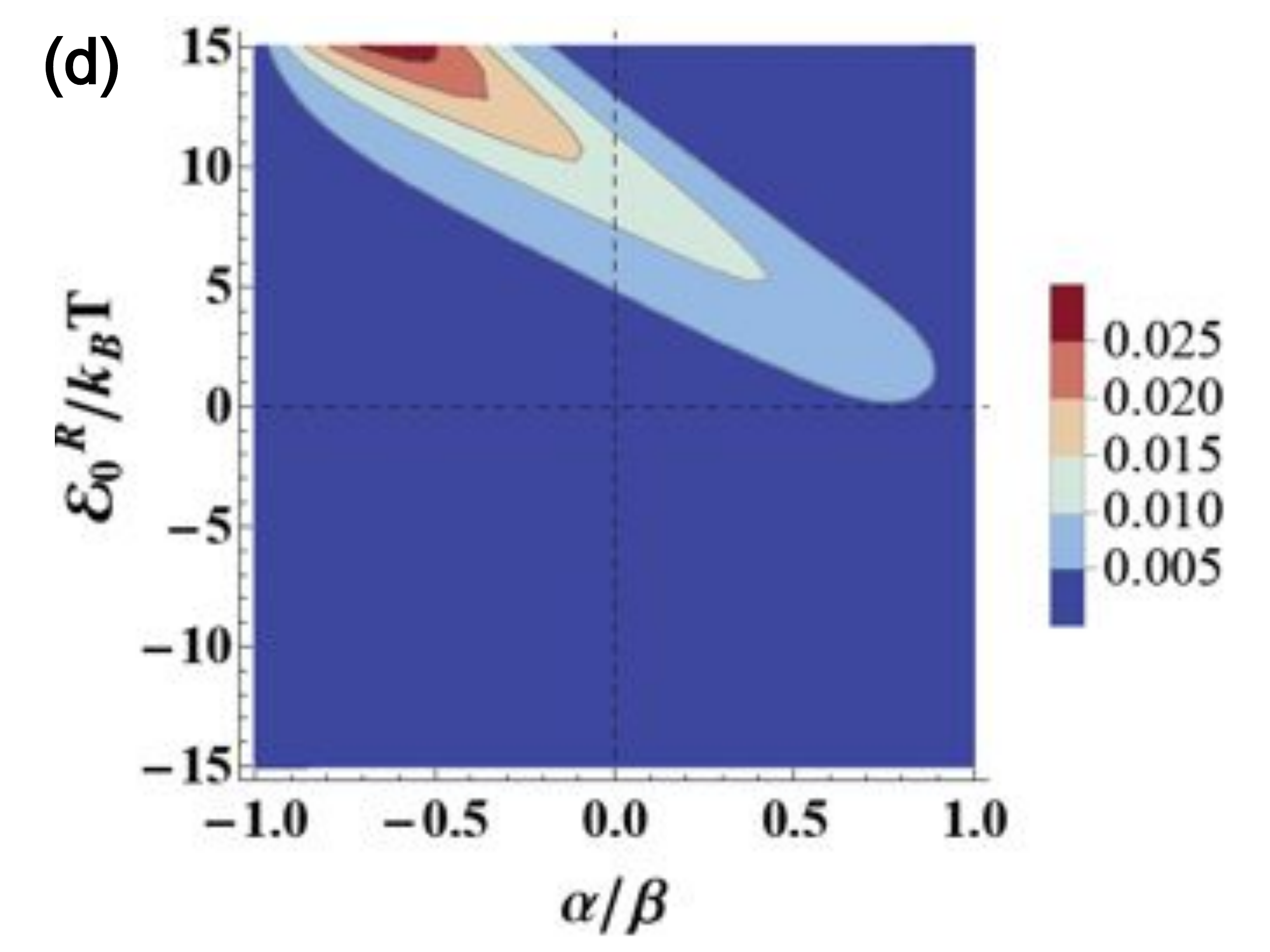}}
\caption{(a), (b) Maximum power in units of $\frac{\nu_{2D}\mathcal{A}(k_BT)}{2\hbar}(\frac{k_B\Delta T}{2})^2$ within linear response as a function of the center of right miniband $\mathcal{E}_{0}^R$ and the asymmetric parameter $\alpha/\beta$, when (a) is $\beta=k_BT$ and $\mathcal{E}_0^L=-4k_BT$, and (b) is $\beta=5k_BT$ and $\mathcal{E}_0^L=-20k_BT$. (c), (d) Efficiency at maximum power normalized by Carnot efficiency $\eta_c$ within linear response as a function of $\mathcal{E}_{0}^R$ and $\alpha/\beta$. Panel (c)/(d) is a corresponding efficiency for the case of (a)/(b). All plots are obtained for $T=300K$ and $\Delta T=1K$.}\label{fig_linear_maxpower_asym}
\end{center}
\end{figure}

Fig. \ref{fig_linear_maxpower} shows the maximum output power and the corresponding efficiency on equal size bands ($\alpha=0$) as a function of the centers of minibands of the two superlattices, $\mathcal{E}_0^{L/R}$. Both the power and efficiency are symmetric with respect to an exchange of $\mathcal{E}_0^L$ and $\mathcal{E}_0^R$. On the one hand, the maximum power arises when one of the two center value stays around twice of $\beta$ and the other center is deep below the equilibrium chemical potential, below about $-3\beta$. The center around $2\beta$ is the position that makes the bottom of the miniband stay around the equilibrium chemical potential, and the other center which is below $-3\beta$ makes the top of the miniband deep below the chemical potential, because the miniband width is $4\beta$. On the other hand, the efficiency acts symmetrically depending on the position of band centers as well as the power. However, maximum efficiency comes when one of the center of miniband is bigger than $2\beta$ with an appropriate value of the other center, and it comes with suppressed output power. Therefore, depending on what we want to optimize, the output power or efficiency, we can chose the position of band centers.

We also show the maximum power for the different miniband widths from the top panels of Fig. \ref{fig_linear_maxpower}.
As the electrons transport only within the miniband, a wider miniband allows more electrons to transport and generate more power. However, as the miniband width goes too far above $k_BT$, the power increase stops. The reason is that the energy window of the order $k_BT_{c/r}$ will be a more effective energy guard for the carriers.
Different from the power, efficiency is reduced as the miniband width increases, Eq. (\ref{fig_linear_maxpower_asym}(c),(d)). As the width increases, the energy filtering by the superlattices is lesser efficient so the efficiency decreases. However, as the width increases continuously, the efficiency will saturate at some point with the same reason for the power saturation.

These results show that the superlattice heat engine has more power output than the quantum well engine \cite{Sothmann2}. The maximal output power of the superlattice heat engine with units $P_0=\nu_{2D}\mathcal{A}k_BT/(2\hbar (k_B\Delta T/2)^2)$ is $P_{max}\approx 1.8 P_0$ with the efficiency $\eta_{maxP}\approx0.05\eta_{c}$ (for the miniband width, $\beta=k_BT$), while the quantum well heat engine is $P_{max}\approx P_0$ with the efficiency $\eta_{maxP}\approx0.07\eta_{c}$. From Eq. (\ref{gr1}-\ref{gr3}), we can give a simple reason for this: when the miniband width is suppressed (longitudinal energy window becomes one value and the electrons transport only with the certain longitudinal energy), our results approach the quantum well case. Therefore, a large miniband width permits more electrons to transport in the longitudinal degree of freedom. While the superlattice engine is more powerful, it is less efficient an energy filter than the quantum well case: this is from the difference between the miniband of the superlattice and the sub-band threshold of the quantum well.
\begin{figure}
\begin{center}
\subfigure{\includegraphics[width=4.2cm]{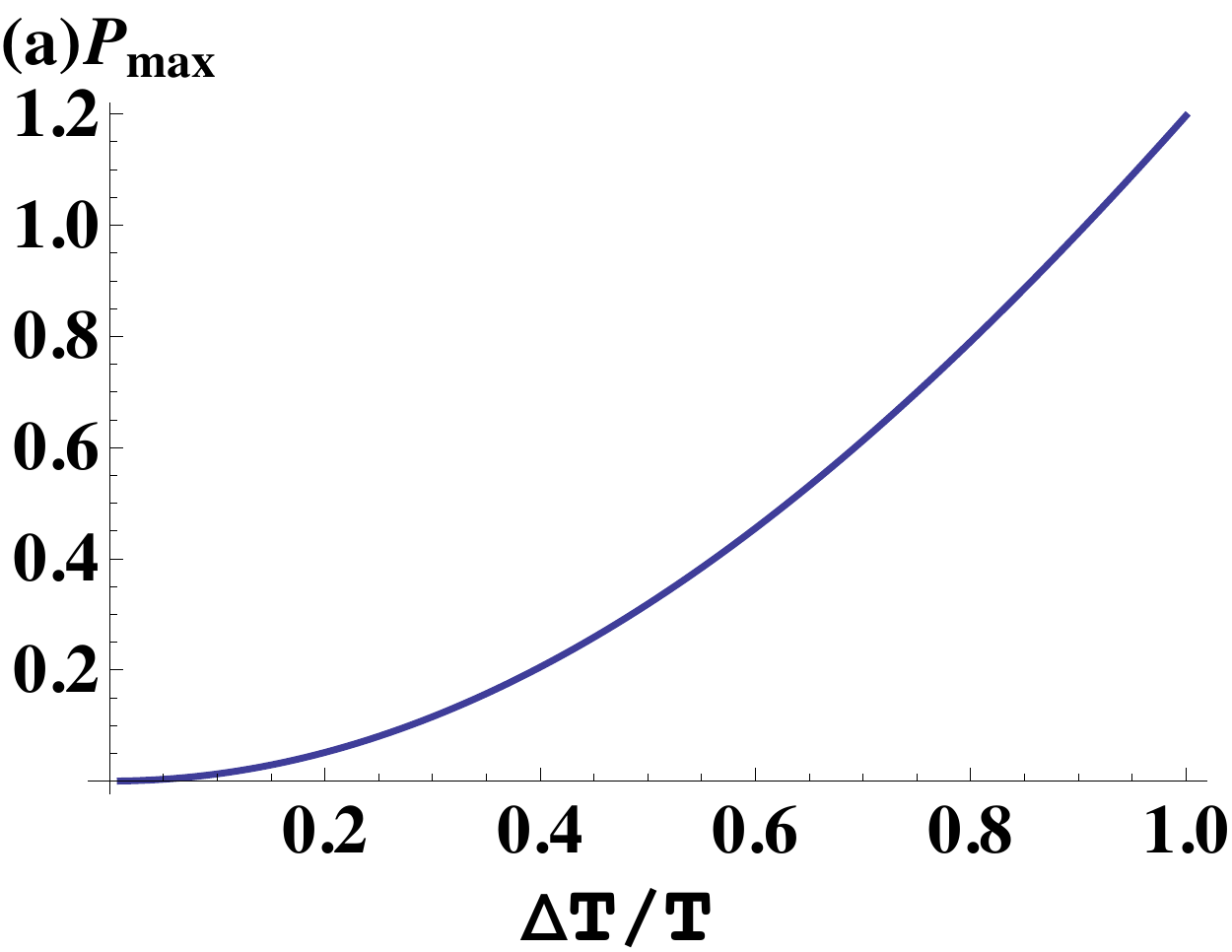}}
\subfigure{\includegraphics[width=4.2cm]{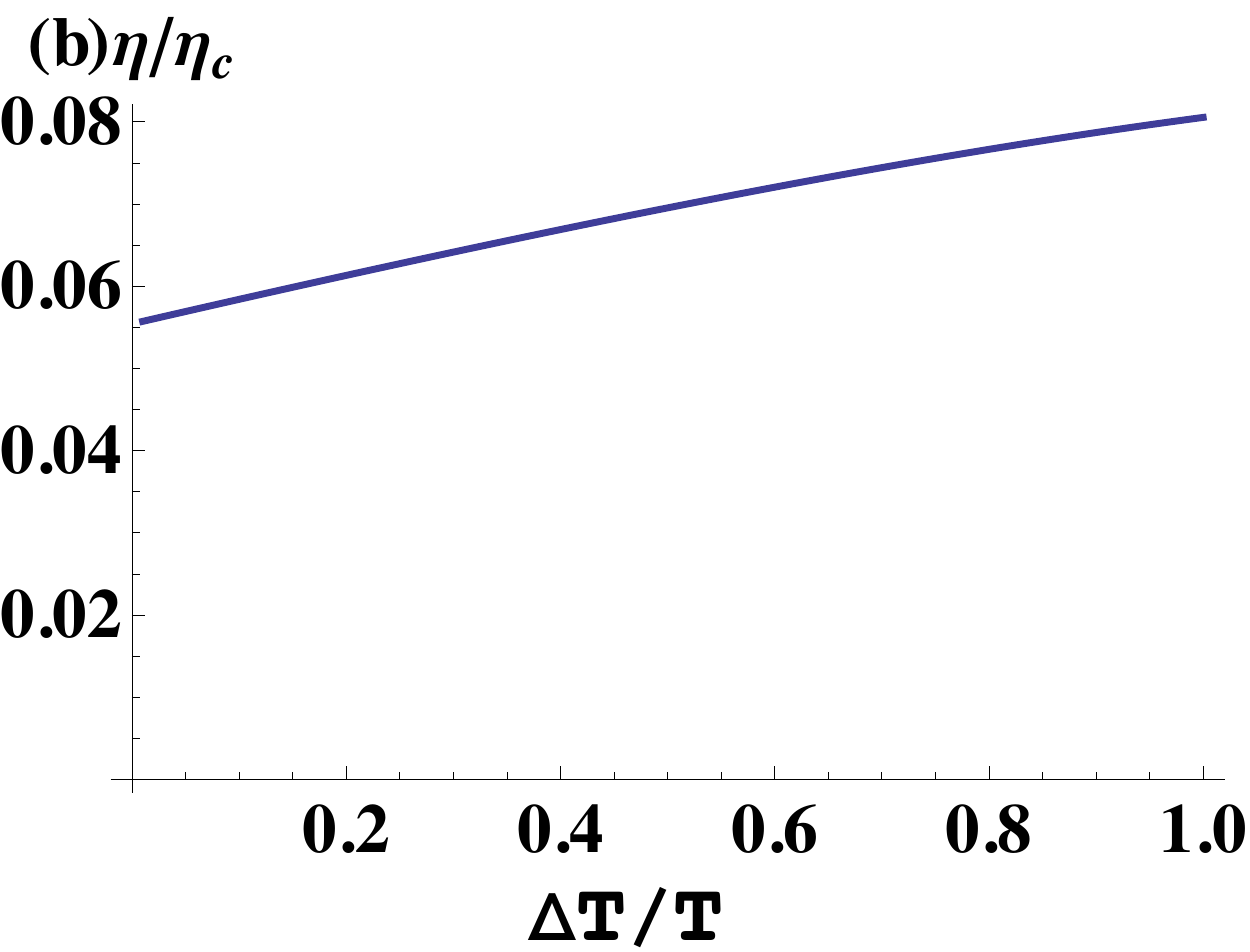}}
\caption{The figure shows nonlinear response when the miniband width is $\beta=k_BT$. (a) the maximum output power as a function of temperature difference $\Delta T/T$ for the optimized parameters $\mathcal{E}_0^L=-5 k_BT$, $\mathcal{E}_0^R=2k_BT$, and $\alpha/\beta=-0.25$. (b) shows the corresponding efficiency. The power is in the units of $\frac{\nu_{2D}\mathcal{A}(k_BT)^3}{2\pi\hbar}$.}\label{fig_non}
\end{center}
\end{figure}

We now turn to an asymmetric system, $\alpha\neq0$. To find the optimized combination of the centers of miniband and the asymmetric parameter $\alpha$ for the maximum output power, we consider the output power and the efficiency for $\mathcal{E}_0^L = -4\beta$. Fig. \ref{fig_linear_maxpower_asym} shows plots of the power as a function of $\alpha$ and $\mathcal{E}_0^R$. When $\beta=k_BT$, the largest output power $P_{max}\approx1.85 P_0$ (around a $2\%$ increases from the symmetric case) arises for $\alpha\approx -0.2\beta$ with $\mathcal{E}_0^R\approx 2.6k_BT$. In this case, the efficiency goes to $\eta_{maxP}\approx 0.06\eta_c$ (a $20\%$ increase from the symmetric case).
Meanwhile, the maximum efficiency $\eta\approx 0.16\eta_c$ comes with $\alpha\rightarrow -\beta$ and $\mathcal{E}_0^R\rightarrow 5k_BT$ for $\beta=k_BT$. When $\alpha\rightarrow -\beta$, the miniband widths become $\beta^L\rightarrow 0$ and $\beta_R\rightarrow 2\beta$, and the output power is strongly suppressed. We can understand the reason for this last fact because when the width of the left miniband vanishes, transport from the left is cut off, giving no power, but very good energy filtering, which increases efficiency.   To explain other features of the plots, we note that for the right superlattice, the center of miniband stays slightly above the equilibrium chemical potential, and the width of miniband $\Delta=4\beta_R$ is larger than the width of the distribution function $\sim k_BT$. Consequently, making the miniband wider has no effect on the electron transport because the number of electrons with energies around minimum/maximum of the energy window is exponentially small. Simultaneously, for the smaller miniband of the left barrier, centering it around the energy region where the left reservoir is occupied but the cavity not, gives very good current production.  As the miniband width $\beta$ increases, Fig.~(\ref{fig_linear_maxpower_asym}b), $\alpha$ moves to a positive value while the center $\mathcal{E}_0^R$ stays around the value that maximizes the power for the symmetric case. Therefore, depending on the size of the miniband width relative to the width of the occupation function of the reservoir and the cavity, $\alpha$ can be fixed to give the optimized results.

\subsubsection{Nonlinear response}
It is interesting next to consider the output power and the efficiency in the nonlinear regime, where qualitatively new physics can appear \cite{DSanchez,Meair,Whitney2013}.   We numerically calculated the stopping voltage $V_{stop}$, the centers of miniband $\mathcal{E}_0^{L/R}$, and asymmetry parameter $\alpha$ in order to maximize the output power.
For the case $\beta=k_BT$, we show the maximum output power and the efficiency in Fig.~\ref{fig_non}. This is obtained for the miniband centers $\mathcal{E}_0^L=-5 k_BT$ and $\mathcal{E}_0^R=2k_BT$ with the asymmetric parameter $\alpha=-0.25\beta$. The output power increases quadratically in the temperature difference for the fixed average temperature $T$ as in the linear case, Eq. (\ref{maxpower_linear}), but it also depends on the average temperature, $P_{max}\sim T\Delta T^2$. Therefore higher average temperature as well as the temperature difference gives bigger output power.

In Table \ref{table_comparison}, we compare the maximum output power and the corresponding efficiency for three systems: the quantum dot \cite{Jordan}, quantum well \cite{Sothmann2}, and superlattice based three terminal heat engines for a realistic device parameter $m_{eff}=0.067m_e$ with the room temperature $T=300K$ and the temperature difference $\Delta T=1K$. We see that $P_{max}$ of the superlattice heat engine generates a larger power about $0.3\textmd{W}/\textmd{cm}^2$ with a small reduction for the efficiency $0.06\eta_c$. Therefore, the superlattice heat engine is the more powerful heat engine.

\begin{table}[t!]
\centering
\begin{tabular}{ |c|c|c|c| }
\hline
 & Quantum Dots & Quantum Wells & Superlattices \\
\hline
\hline
$P_{max}$($\textmd{W}/\textmd{cm}^2$) & 0.1 & 0.18 & 0.3 \\
\hline
$\eta_{maxP}$($\eta_c$)& 0.2 & 0.07 & 0.06 \\
\hline
\end{tabular}
\caption{We compare our result with other three terminal geometries: (power-optimized resonant width) quantum dot and quantum well heat engines for $T=300K$ and $\Delta T=1K$. }\label{table_comparison}
\end{table}

\subsection{Quantum refrigerator based on superlattices}\label{refrigeration}
Now, we consider the same geometry but for a different purpose, a cooling system. If proper bias voltage $eV$ is applied over the junction and the minibands of superlattices are suitably arranged, a current flows from right to left as hot electrons tunnel through the miniband of the left superlattice from the cavity to left reservoir and cold electrons from the right reservoir tunnel through the right miniband to the cavity.
This leads to decrease in the average energy of electrons in the central cavity, that is cooling utilizing the Peltier effect: in the Peltier effect, the junction is electrically biased and a produced heat current flow given an electric current.

For the refrigerator, $\Delta T$ is negative in our notation, and the temperature of the center cavity cooled down to the amount of $|\Delta T|$.
The purpose of refrigeration is to achieve a large temperature reduction. The base temperature of the refrigerator is defined as the temperature for equilibrium where the evacuated heat current balances any external heat leaks. Since we are now considering no external heat leaks, such as electron-phonon coupling, the base temperature is the temperature for which $J=0$ in Eq. (\ref{heatcurrent_linear}) with the temperature reduction
\begin{eqnarray}
\Delta T_{0}=-\frac{G_e\Pi}{G_eS_t\Pi+H_{t1}+H_{t2}}V,\label{base_temp}
\end{eqnarray}
in the linear regime. The refrigeration works only when the heat current is emitted by the cavity into the reservoirs, $J\geq 0$, therefore the temperature reduction of the cavity should be in the range of $\Delta T_{0}<\Delta T<0$.

The applied bias voltage $V$ and the absorbed heat from the cold cavity let the heat flow $J$ rejected to the left and right reservoirs. Therefore, the cooling power is $J$, and an efficiency of the cooling is normally characterized by the coefficient of performance (COP), defined as the ratio of the cooling power to the total input power $P=IV$,
\begin{eqnarray}
\phi=\frac{J}{P}.
\end{eqnarray}
Similar to the efficiency of the energy harvester, COP is also bounded by the Carnot value, $\phi\leq\frac{T_c}{T_r-T_c}=\phi_{c}$.

\begin{figure}[t]
\begin{center}
\subfigure{\includegraphics[width=8cm]{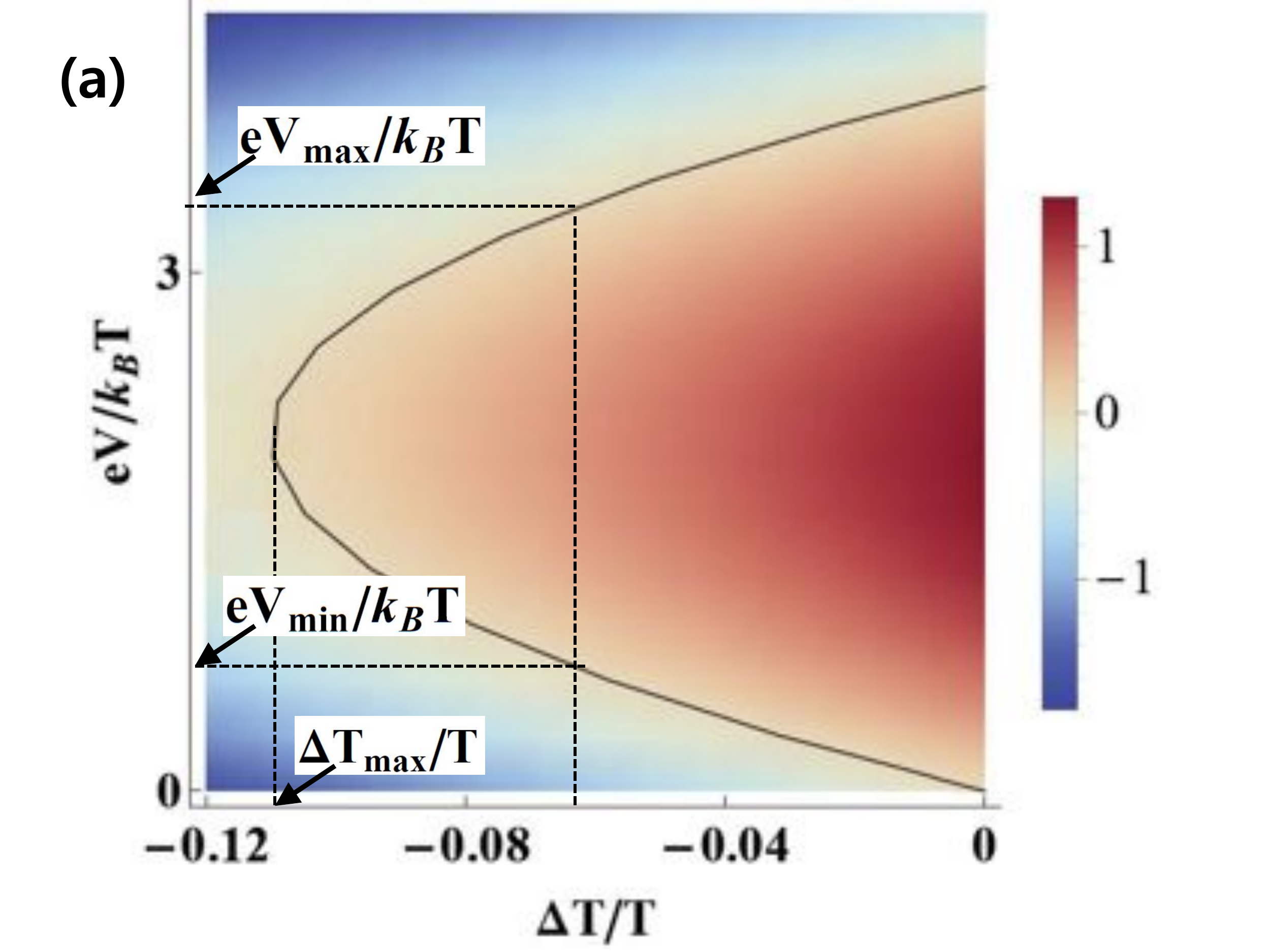}}
\subfigure{\includegraphics[width=8cm]{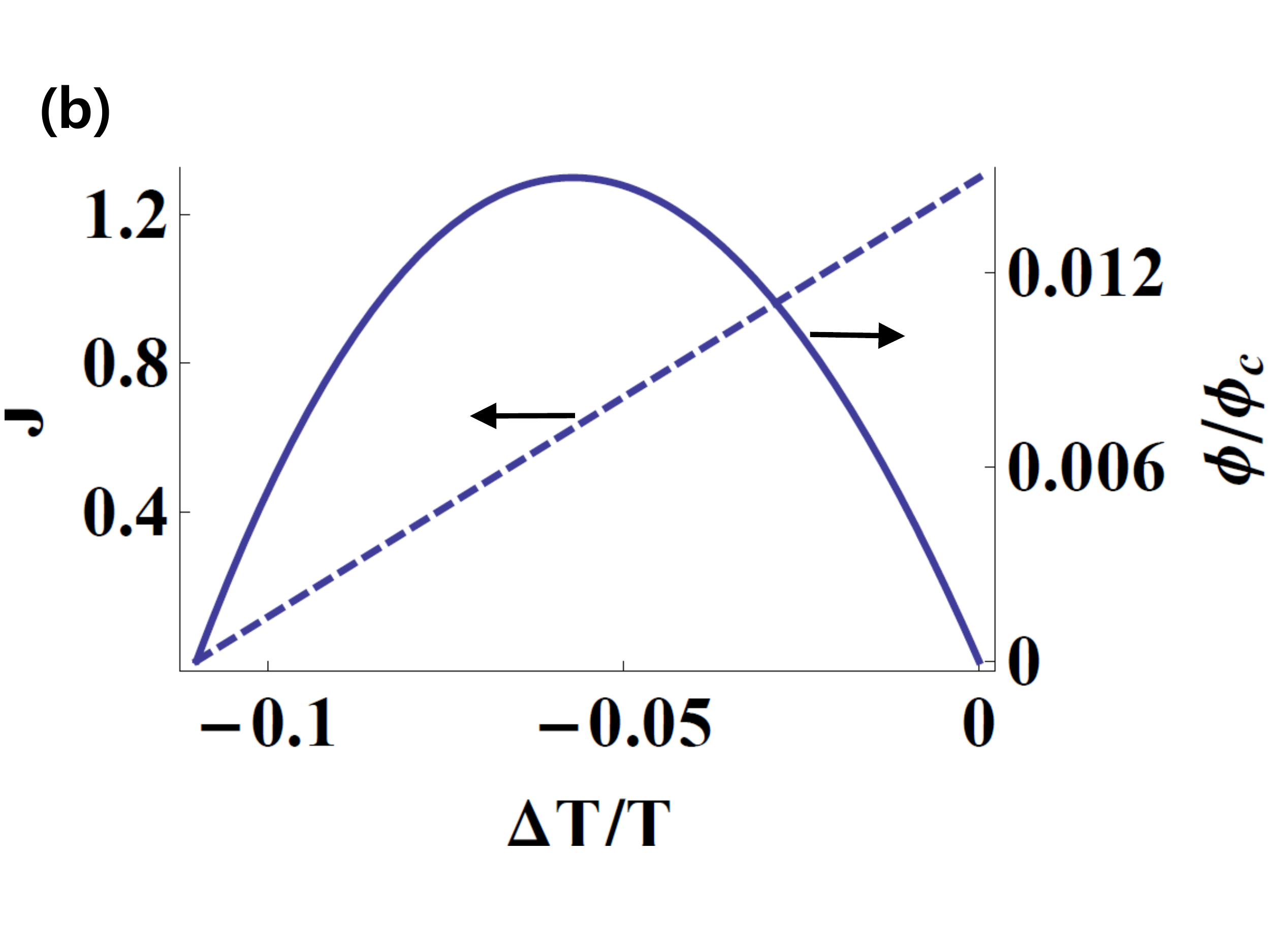}}
\caption{(a) The nonlinear cooling power is shown as a function of the temperature difference and the bias voltage when the other parameters, $\mathcal{E}_0^L$, $\mathcal{E}_0^R$, $\beta$, and $\alpha$, are optimized to give the maximum cooling power. 
The cooling power is in the units of $\frac{\nu_{2D}\mathcal{A}(k_BT)^3}{2\pi\hbar}$.  Panel (b) shows the heat current and the COP as a function of the temperature difference when the bias is chosen to give the maximum cooling power, $eV\simeq 2k_BT$. }
\label{fig_rfg}
\end{center}
\end{figure}

Moving beyond the linear regime is important to find the nonlinear behavior of the refrigerator, which determines the lowest temperature it can reach.  Fig. \ref{fig_rfg}(a) shows a plot of the cooling power vs applied bias and temperature differences, when other variables, $\mathcal{E}_0^R=-5k_BT$, $\mathcal{E}_0^L=2k_BT$, $\beta=k_BT$ and $\alpha=0.3\beta$, are optimized to give the largest cooling power for the positive applied bias $V>0$. Here, $\alpha>0$ means that left miniband width is wider than the right miniband, and the reason can be understood as the energy harvesting that we discussed in section \ref{result}. Note that, the position of the left/right miniband of the refrigerator is opposite from the energy harvesting, so this makes the conduction electrons take energy from the cavity and continue to cool it.

The black curve in panel (a) of Fig.~(\ref{fig_rfg}) represents when the cooling power becomes zero, and the right hand side of the line (within the parabola) is for the positive cooling power which is the region that works for the refrigeration. Therefore, we can see the optimal region for the temperature difference and the applied voltage, as shown in the plot.
When the temperature difference is anywhere between the zero and $\Delta T_{max}$, as an example the vertical line in the figure, the applied voltage should stay in between $V_{min}$ and $V_{max}$ to make the engine work as the refrigerator.
Specially, when the applied voltage is around $2k_BT$, the refrigerator gives maximum cooling power for a given temperature difference, and we also can have the maxumum temperature reduction $|\Delta T_0|$ of the system. When $|\Delta T|\rightarrow 0$, the heat current increases linearly in panel (b). So, the maximum temperature reduction for this bias voltage is $|\Delta T_{0}|=|\Delta T_{max}|\simeq 0.11 T$. For example, at room temperature, the maximum temperature reduction is $|\Delta T_{max}|=30K$ by applying the bias around $50 \textmd{meV}$. If the temperature reduction is $|\Delta T_{0}|=0.05 T=15K$, the corresponding cooling power is approximately $11\textmd{kW/cm}^2$.
The COP at $eV=2k_BT$ as a function of the temperature difference is also plotted in Fig. (\ref{fig_rfg}(b)). The maximum COP $\phi\simeq 0.015\phi_c$ comes with when $|\Delta T|\simeq 0.06 T$. Therefore, the refrigerator can be operated at the optimal regime for the cooling power or COP by reasonably choosing the parameters.

\section{Heat transport by phonons}\label{heat_engine_phonon}
\subsection{Reduced phonon heat current}
One important consideration in thermoelectric heat engine is the heat flow carried not only by the conduction electrons, but by the phonons as well. Here, we calculate this effect for our system to see how it will affect the efficiency in a more realistic modeling.

The major heat flow from hot to cold reservoirs is carried by excitations such as phonons. For this three terminal heat engine, the phonon heat flow $J^{ph}$ is in parallel with electronic flow $J^{e}$, and the total heat flow for a given generated power $P$ is the sum of them, $J=J^{e}+J^{ph}$. Then we can rewrite the efficiency in the present of the phonons
\begin{eqnarray}
\eta^{e+ph}=\frac{P}{J^e+J^{ph}}.
\end{eqnarray}

\begin{figure}[t!]
\begin{center}
\subfigure{\includegraphics[width=4.2cm]{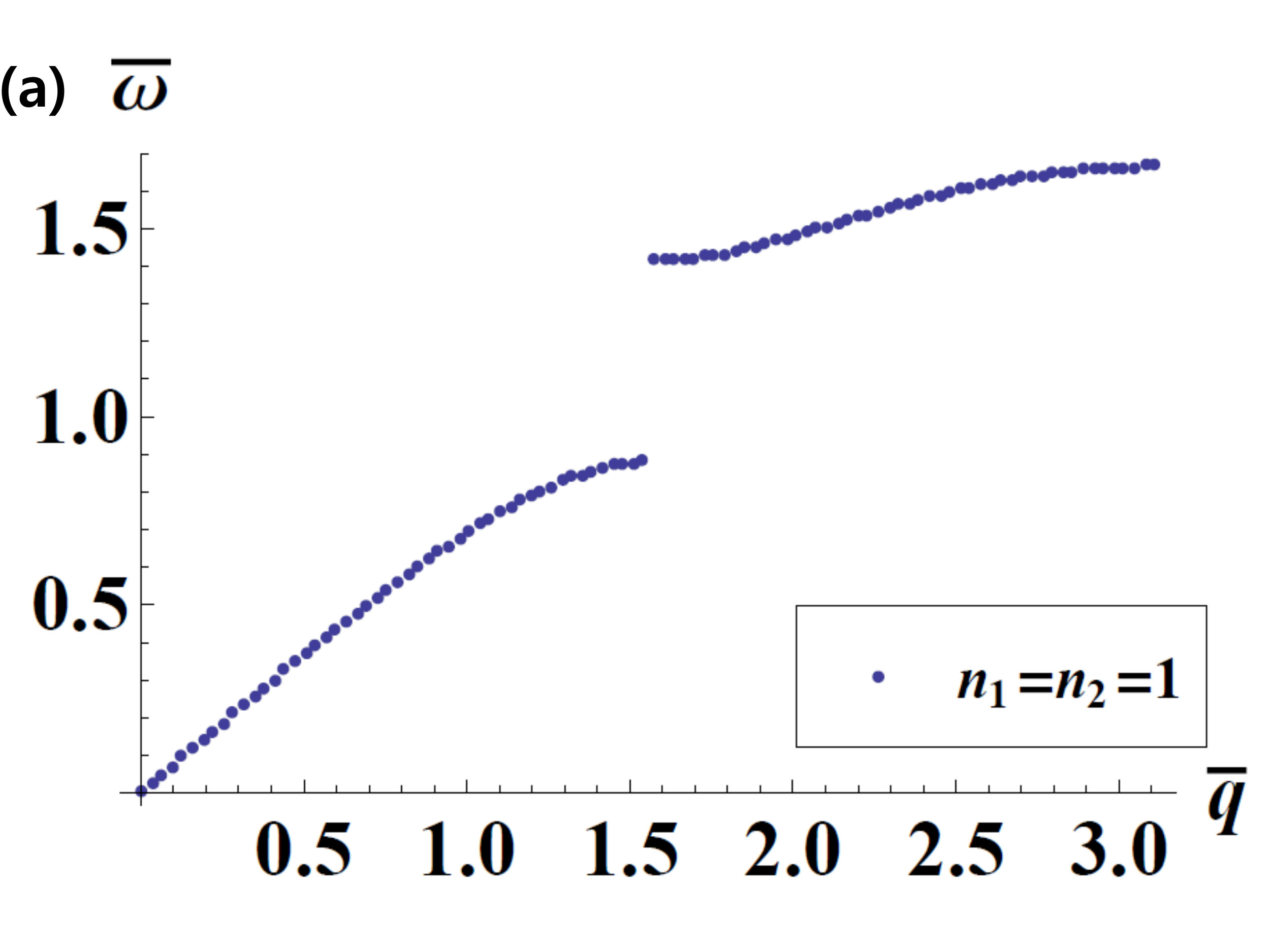}}
\subfigure{\includegraphics[width=4.2cm]{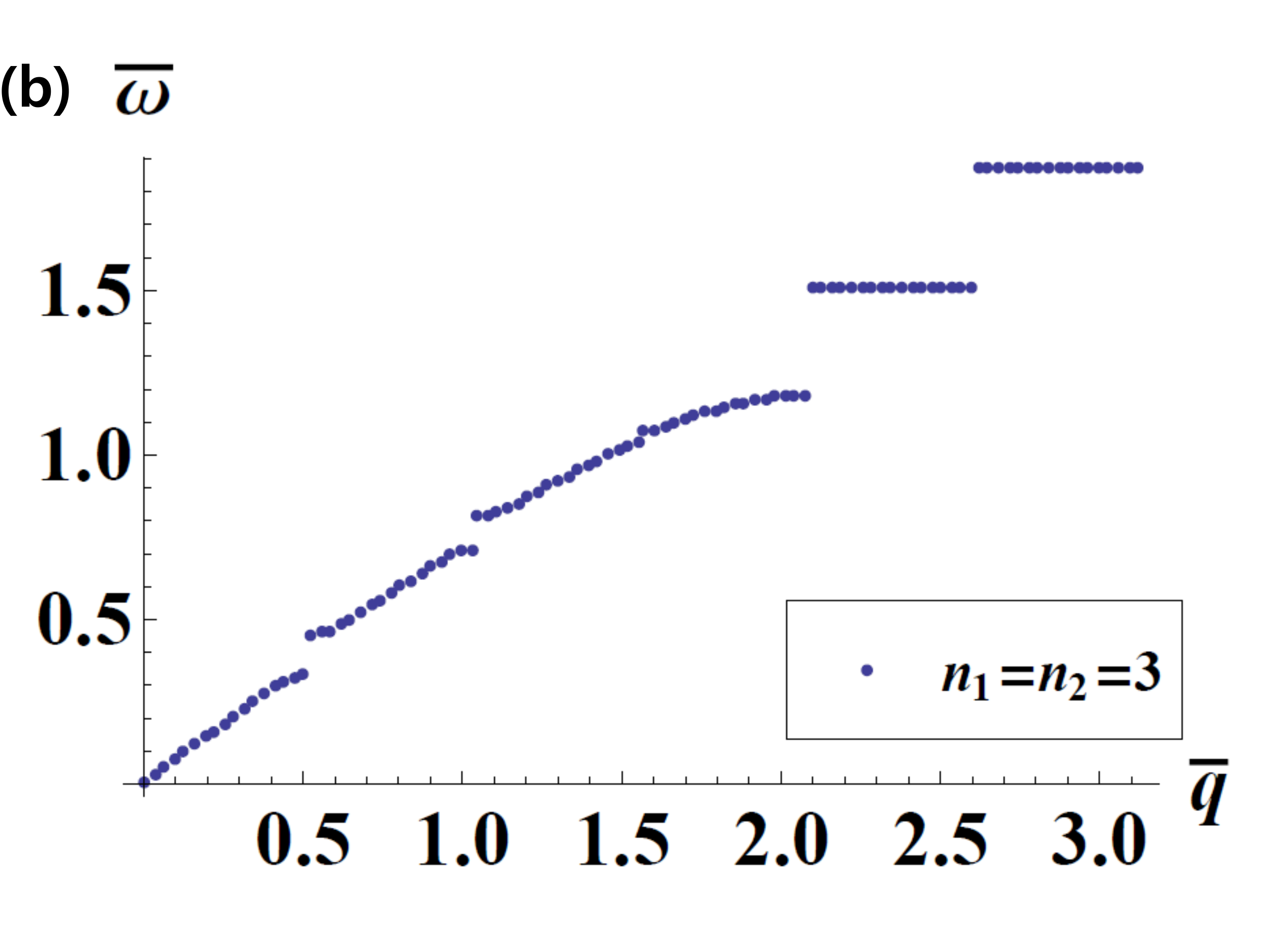}}
\caption{The phonon dispersion curves from a one dimensional atomic chain for the case of mass ratio $\delta=2.6$ in the extended zone representation. (a) and (b) show $n_1=n_2=1$ and $n_1=n_2=3$, respectively. The dimensionless parameters are defined as $\overline{q}\equiv qa$ and $\overline{\omega}\equiv \omega\sqrt{M_1/g}$.}\label{fig_phonon_disp}
\end{center}
\end{figure}

The phonon heat current in the superlattice differs from the heat current based on the bulk material properties due to the new periodicity of the structure. The presence of interfaces can alter the phonon spectra result from wave interference scattered at the interfaces. The formation of miniband gaps in superlattices due to phonon interference leads to a reduction of the phonon velocity $v_{g}(k)$ which gives a reduction of the thermal conductivity.
To illustrate this effect in a simple model, we assume the complete separation of longitudinal and transverse vibrations for phonons with the wave vector to parallel to $z$ axis. Also assuming perfect interfaces, the transverse momentum $q_{x(y)}$ is conserved. When we consider phonons propagating in the growth direction, this can be treated from a Kronig-Penney type approach for one dimensional atomic chain \cite{Colvard}. We suppose the same monolayer spacings $a$ and the magnitude of atomic constants $g$ in between all the atoms. The layer one(two) has $n_1(n_2)$ atoms with mass $M_{1}(M_{2}=\delta M_{1})$, and the thickness of sublattice one (two) is $d_1=n_1 a$ ($d_2=n_2 a$) which gives the length of unit period $d=d_1+d_2$. Then the characteristic equation is
\begin{eqnarray}
\lefteqn{\cos(q_z d)=\cos(k_1d_1)\cos(k_2d_2)}\nonumber\\
&-&\frac{1-\cos(k_1a)\cos(k_2a)}{\sin(k_1a)\sin(k_2a)}\sin(k_1d_1)\sin(k_2d_2), \label{characteristic equation}
\end{eqnarray}
where $k_{1(2)}$ is the phonon wave vector of each layer $1(2)$ with $\cos(k_{1(2)}a)=1-M_{1(2)}\omega^2/2g$, and $q_z$ is a longitudinal superlattice wave vector. This model gives us the dispersion of the longitudinally polarized phonons for the cross plane transport with zero transverse momentum. The dispersions for different periods are shown in Fig. \ref{fig_phonon_disp} for $n_1=n_2=(1,3)$ with the mass ratio $\delta=2.6$. Increasing the superlattice periods ($n_1$ and $n_2$) gives more band folding and decreases the average group velocity.

At nonzero transverse momentum, the longitudinally polarized phonon mode for the cross plane dynamics is described as \cite{Hyldgaard}
\begin{eqnarray}
\omega_{\textbf{q}}^2=\omega_{q_z}^2+\Omega_{t}^2\frac{1}{2}(2-\cos(q_x a)-\cos(q_y a)),
\end{eqnarray}
where $\omega_{q_z}$ is a solution from Eq. (\ref{characteristic equation}) and $\Omega_{t}$ is a characteristic frequency of the material. Therefore, the dispersion of non zero transverse momentum can be used to calculate the group velocity in the superlattices, $\vec{v}_g=\nabla_{\vec{q}}\:\omega_{\textbf{q}}$. Since we are dealing with the heat transfer only to the cross plane direction, the group velocity that we need is only the $z$ direction,
\begin{eqnarray}
v_{g,z}=\frac{\omega_{q_z}}{\omega_{\textbf{q}}}v_{g,1d},\label{group_vel}
\end{eqnarray}
where we use the one dimensional group velocity $v_{g,1d}\equiv \partial\omega_{q_z}/\partial q_{z}$. Now, we can write the phonon heat current from the reservoir $r$ into the cavity through the superlattice,
\begin{eqnarray}
J_r^{ph}=\int\frac{d^3\vec{q}}{(2\pi)^3}\:\hbar\omega_{\textbf{q}}\:v_{g,z}(\vec{q})\:\mathcal{T}(q_z)\:\Delta B(\vec{q},T_r,T_c),\label{phonon_heat_current_mom_gen}
\end{eqnarray}
where $\Delta B(\vec{q},T_r,T_c)=b_r(\vec{q})-b_c(\vec{q})$ is the occupation difference between the reservoir and cavity with the Bose-Einstein distribution function $b_{r/c}(\vec{q})=(\textmd{exp}(\hbar\omega(\vec{q})/k_BT_{r/c})-1)^{-1}$ and $\mathcal{T}(q_z)$ represents the transmission function which depends only on the longitudinal momentum $q_z$. In our analysis, we assume the ballistic case with $\mathcal{T}(q_z)\simeq1$. When we put the longitudinal group velocity Eq.(\ref{group_vel}) into the first equation of Eq.~(\ref{phonon_heat_current_mom_gen}), we can factorize the heat current to the longitudinal $(z)$ and transverse $(x,y)$ directions
\begin{eqnarray}
J_r^{ph}=\int_{0}^{\pi/d} \frac{dq_z}{2\pi}\:\hbar\omega_{q_z}\: v_{g,1d}(q_z)\:\Delta F(q_z,T_r,T_c),\label{phonon_heat_current_mom}
\end{eqnarray}
with a modified occupation difference
\begin{eqnarray}
\Delta F(q_z,T_r,T_c)=\int_{-\pi/a}^{\pi/a}\frac{dq_xdq_y}{(2\pi)^2}\:\Delta B(\vec{q},T_r,T_c).
\end{eqnarray}
This function gives the effective phonon occupation difference per unit area between the cavity and reservoir due to the transverse momentum. The transverse momentum of phonons contributes only through the distribution function, and the energy and velocity dependencies of the heat current are only from the longitudinal momentum $q_z$. This analysis suggests that the heat current can be treated as a one dimensional calculation with an effective occupation function. We rewrite the new expression in terms of an integral over longitudinal frequency instead of the momentum space
\begin{eqnarray}
J_r^{ph}&=&\sum_{i=1}^{n_1+n_2}\int_{\omega_i^{-}}^{\omega_i^{+}}d\omega_{q_z}\frac{v_{g,1d}}{2}D_{1D}(\omega_{q_z})\hbar\omega_{q_z}
\Delta F(\omega_{q_z},T_r,T_c)\nonumber\\
&=&\sum_{i=1}^{n_1+n_2}\int_{\omega_i^{-}}^{\omega_i^{+}}\frac{d\omega_{q_z}}{2\pi}\hbar\omega_{q_z}\Delta F(\omega_{q_z},T_r,T_c),\label{phonon_heat_current_freq}
\end{eqnarray}
where $\omega_i^{-(+)}$ is the minimum (maximum) frequency of the $i^{\textmd{th}}$ miniband of longitudinal momentum from Eq. (\ref{characteristic equation}). The conversion to frequency space introduces the phonon density of state of the superlattice, and we have $D_{1D}(\omega_{q_z})=(v_{g,1d}\pi)^{-1}$. Therefore, the group velocity and the density of states cancel out and we have a simple equation of the phonon heat current in Eq. (\ref{phonon_heat_current_freq}).

\begin{figure}[t]
\begin{center}
\subfigure{\includegraphics[width=7cm]{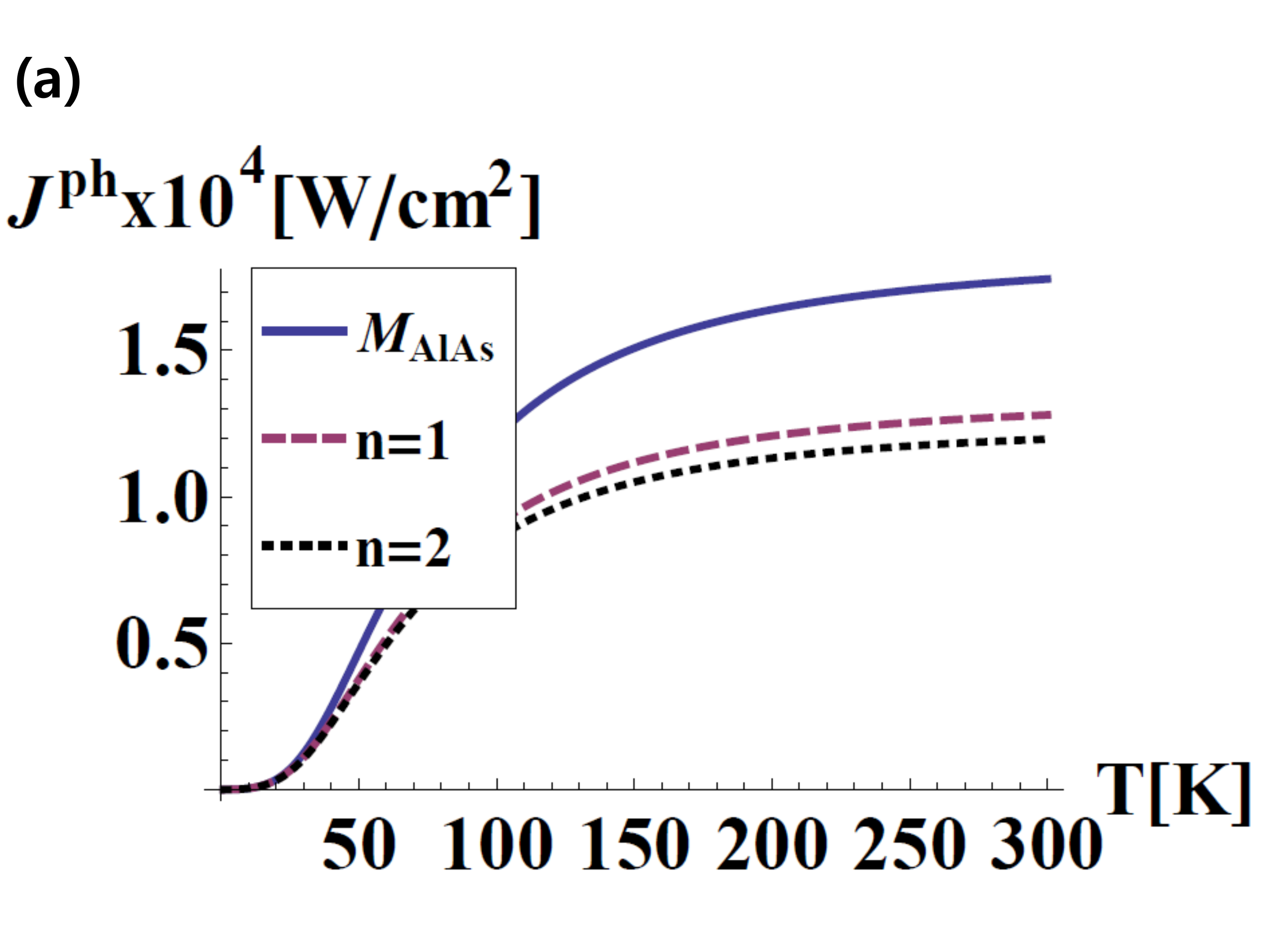}}
\subfigure{\includegraphics[width=7cm]{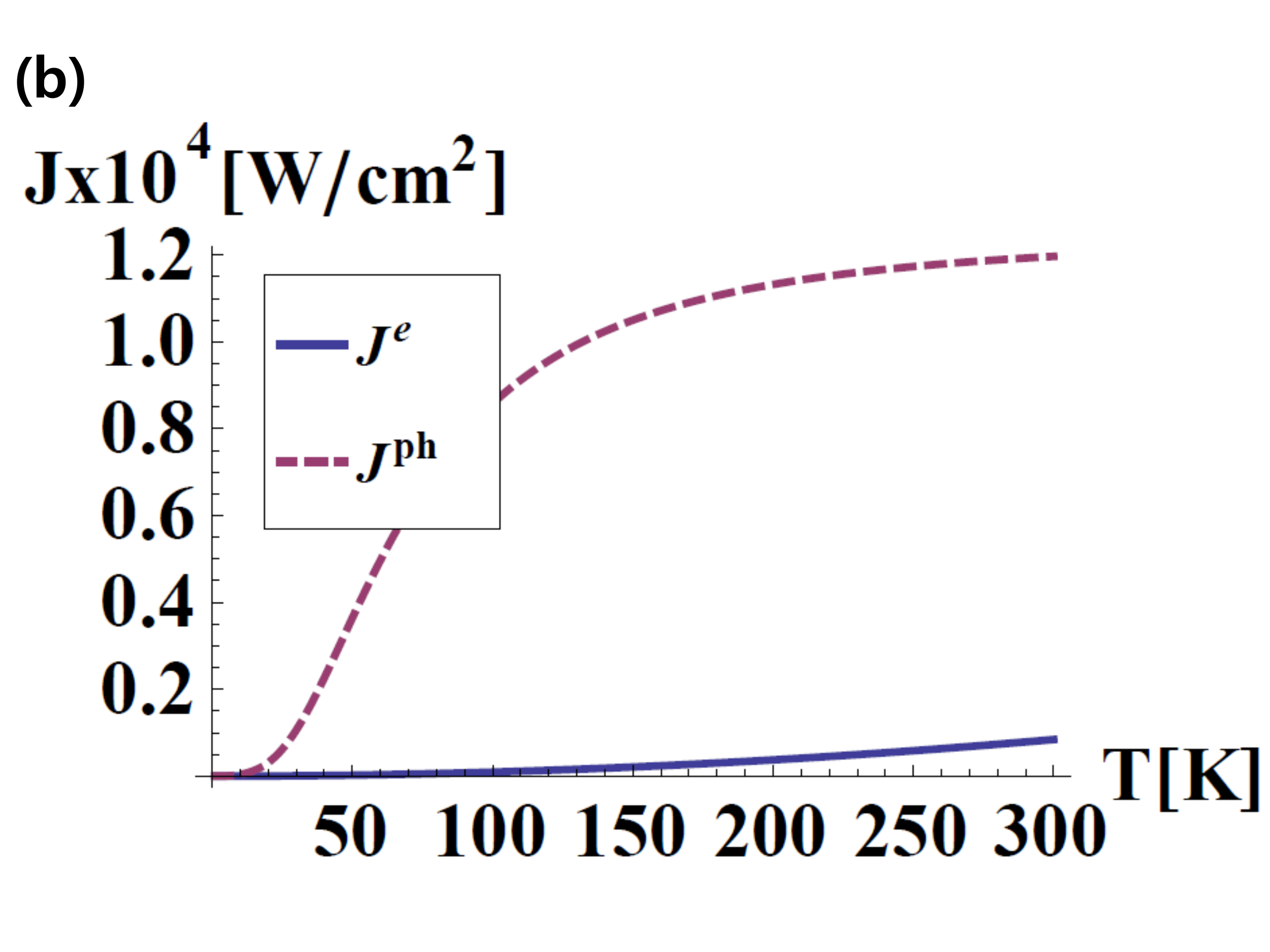}}
\caption{The plots show electron and phonon heat currents of GaAs/AlAs superlattice for the case of the maximum output power with optimized parameters with $\Delta T=1K$. (a) shows the heat current by phonons $J^{ph}$ when the atomic layers are $n_1=n_2= (1, 2)$, with the bulk AlAs case. (b) shows the magnitude comparison between $J^e$ and $J^{ph}$.}\label{fig_phonon_heatcurrent}
\end{center}
\end{figure}

When we consider for the electron transport of the superlattice, note that only the lowest miniband was considered, because the energy gap of the first and second miniband is larger than the thermal energy $k_BT$, for the typical materials of the superlattice. So, when the chemical potentials of the reservoir and the cavity stay around the first miniband, energy window $k_B T$ of the occupation will guide the transport only through the first miniband. Therefore, it is a reasonable approximation to ignore the higher minibands. In contrast, the Bose-Einstein distribution function in the frequency domain is broader than the maximum frequency of the acoustic dispersion, and so all of the acoustic phonons contribute to the heat transport.

The total phonon heat current from the heat engine is also obtained from the heat conservation $J^{ph}+J^{ph}_L+J^{ph}_R=0$. However, different from the heat current by the electrons, the difference of the phonon heat current through the left/right reservoir is only determined by the temperature difference between the cavity and reservoirs which is contained only in the modified occupation difference $\Delta F(\omega_{q_z},T_r,T_c)$. The temperatures of the left/right reservoir are the same in our case, therefore, the phonon heat current conservation leads to the total phonon heat current such as $J^{ph}=-2 J_R^{ph}$ only if there is symmetric heat conductance.


\begin{figure}
\begin{center}
\subfigure{\includegraphics[width=7cm]{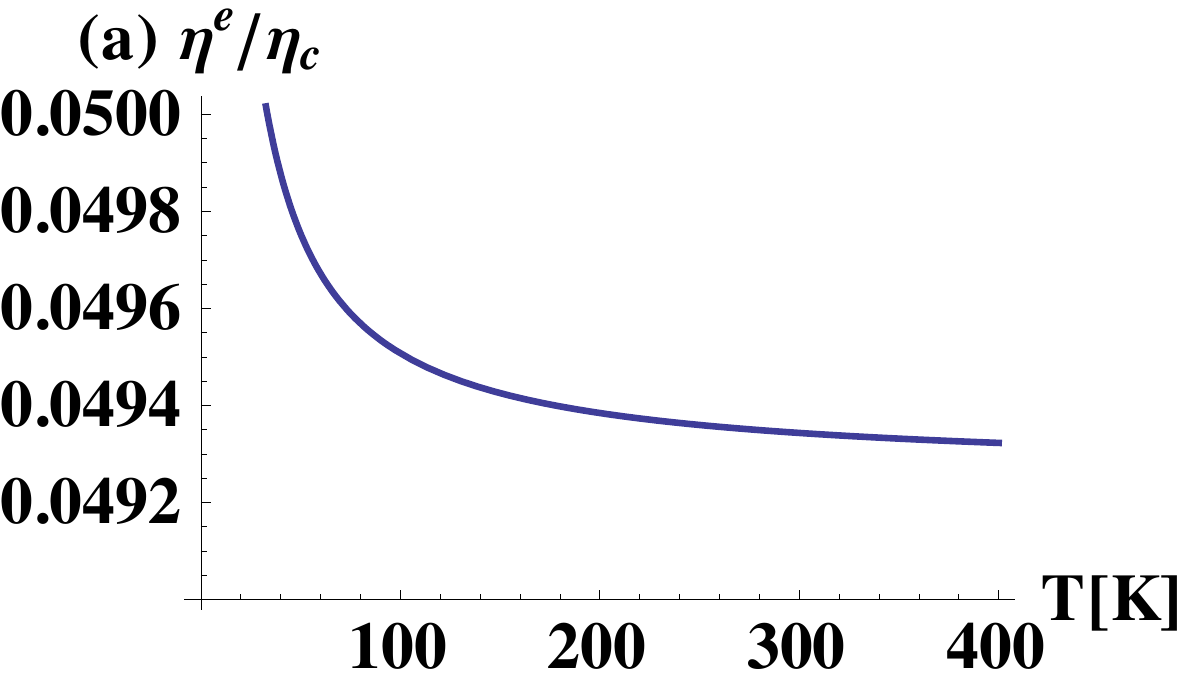}}
\subfigure{\includegraphics[width=7cm]{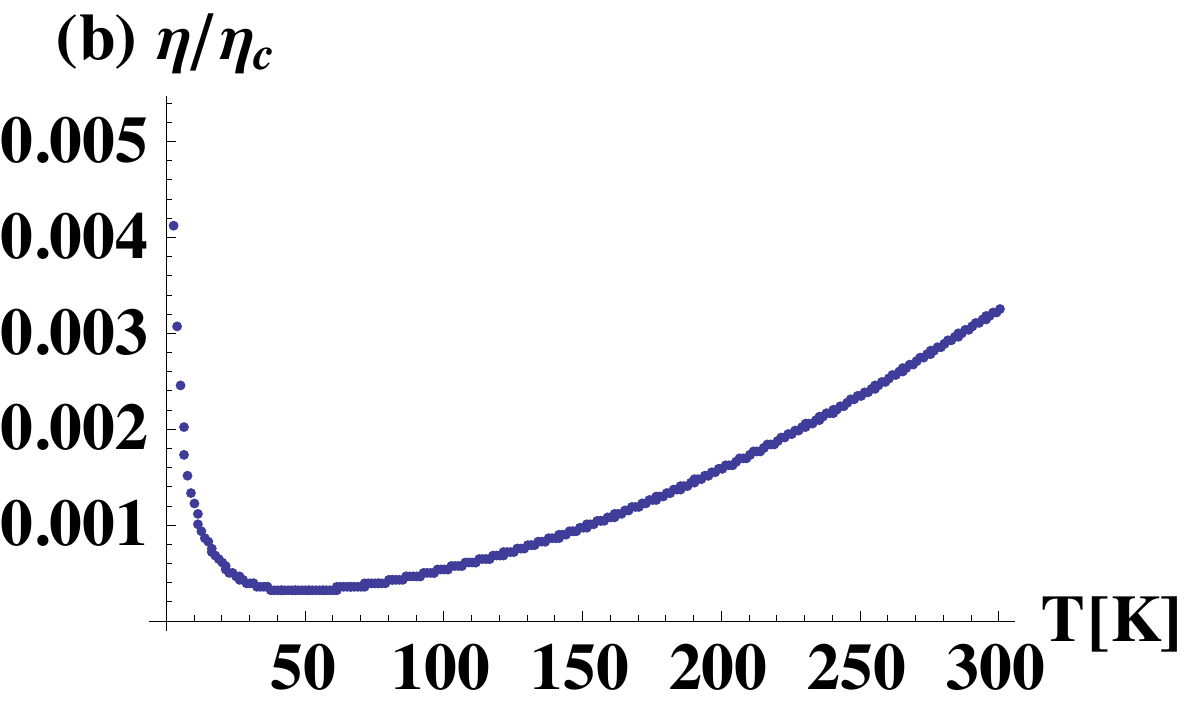}}
\caption{Efficiency at maximum power of GaAs/AlAs superlattices for the value of $\Delta T=1K$, $\beta=k_BT$, and $n_1=n_2=2$, and $\mathcal{E}_{0}^{R/L}$ are chosen to maximize power. The efficiencies by the electron heat current (a) and the total heat current (b) are shown as a function of average temperature.   }\label{fig_total_efficiency}
\end{center}
\end{figure}

In order to calculate the effect of the phonons on the thermodynamic efficiency of this engine, we consider a low temperature regime, where an analytic investigation can be made, followed by a numerical investigation of the high temperature regime.

In the low temperature case, we assume the temperature only excites low energy modes.  Up to a small numerical factor, this permits us to approximate the Bose-Einstein as a Boltzmann distribution, as well as only consider the long wavelength modes, so the dispersion relation may be approximated as, $\omega \approx \sqrt{\omega_{qz}^2 + a^2 \Omega_t^2 (q_x^2+q_y^2)/4}$.
 Computing the effective distribution by integrating over $q_x, q_y$, we find,
\begin{equation}
F(\omega_{qz}) \approx \frac{\hbar \omega_{qz} k_B T}{2 \pi a^2 (\hbar \Omega_t/4)^2} e^{-\hbar \omega_{qz}/k_B T }.
\end{equation}
This allows us to approximate the heat current per unit area from the first mini-band as,
\begin{eqnarray}
J^{ph} &\approx&  \frac{4}{(\pi a \hbar \Omega_t)^2 \hbar}  [ (k_B T_L)^4 - (k_B T_c)^4 ] \\
&\approx& \frac{16}{(\pi a \hbar \Omega_t)^2 \hbar}  (k_B T)^3 \Delta T,
\end{eqnarray}
where the last limit is in the linear response limit.  This is consistent with a Debye treatment of the phonon transport.  This result shows that for low temperatures, the phonons freeze out, and the energy is predominately carried by the electrons.

Moving on to the high temperature limit, we consider a numerical investigation.  This limit is quite important for room temperature applications in mind.  Fig. \ref{fig_phonon_heatcurrent} shows the heat currents for GaAs/AlAs superlattices when $\Delta T=1K$, as an example. The phonon heat currents of superlattices and bulk material in panel (a) increase with temperature increasing, but saturate at high temperature. Moreover, as the superlattice atomic layer increases, the phonon heat current decreases and saturates when the atomic layers are $n_1,n_2 > 10$ \cite{Tamura2}. The practical superlattices usually have the atomic layers $n_1,n_2 > 10$, so, the practical phonon heat current of the superlattices has about $40\%$ reduction of the heat current from the bulk material around the room temperature. Panel (b) compares the electron and phonon heat currents for the optimized parameters to give the maximum output power. As the temperature increases, the electron heat current increases quadratically while the phonon heat current saturates, so the electron heat current reaches the phonon heat current at high temperature. However, at room temperature, the phonon heat current is still an order of magnitude higher than the electron heat current.

We compare the efficiency by the electron heat current and the total heat current in Fig. \ref{fig_total_efficiency}. The phonon heat current dominates over the electronic one, $J^{ph}\gg J^{e}$, when the temperature is around room temperature. In this situation, we can rewrite the total efficiency
\begin{eqnarray}
\eta^{e+ph}\simeq P/J^{ph}.
\end{eqnarray}
The efficiency depends only on the phonon heat current and the generated power. Therefore, we have the maximal efficiency when the power is maximal, and the power and efficiency reach maximal value together. Because the phonon heat current is an order of magnitude higher than the electron heat current, the total efficiency decreases about an order of magnitude.


\subsection{Optimized condition for minimum phonon heat current}
Practically, there could be a way to reduce the phonon heat current more than we showed based on the simple theory and example.
Based on the one dimensional atomic chain model, we have calculated the phonon heat current in the growth direction and estimated their contributions to the efficiency in the heat engine based on GaAs/AlAs superlattices. Our approach shows that the reduction of the phonon heat current comes from the reduction of the group velocity near the folded Brillouin zone edges. This calculation also assumes the ballistic transport $\mathcal{T}(q_z)\simeq 1$. However, as the constituent layers become thicker than the phonon mean free path, the phonon transmission function needs to be modified as $\mathcal{T}(q_z)< 1$. In this case, we also need to treat the phonons as particles and use the theory such as the Boltzmann transport equation \cite{Chen}.
Moreover, other mechanisms such as phonon spectra mismatch and scattering arising from the roughness of the layer interfaces also play an important role to understand the reduction of the experimental results \cite{Chen2,Cahill,Huang1,Huang1,Ren,Tamura2}.

Some of high figure of merit thermoelectric materials show reduced lattice thermal conductivity in the superlattice structures, for example Si/Ge or $\textmd{Bi}_2\textmd{Te}_3/\textmd{Sb}_2\textmd{Te}_3$. The reduction of phonon heat current can be maximized by the proper choice of superlattice period compared to the mean free path of the phonons: a theory predicts the thermal conductivity minimum as a function of layer spacing  \cite{Simkin}, and some works show the minimum thermal conductivity depending on the superlattice period and a ratio of the layer thickness for the materials \cite{Venkat1,Venkat2,Liu}. Therefore, these materials, instead of our example, with optimized superlattice period will give more reduction of phonon heat current of the system.

\section{Conclusions}\label{conclusion}
In this paper, we investigated a heat engine by thermoelectric effects in superlattice structures in three terminal geometry. First, our work considers the engine as a energy harvester, and shows the advantages of superlattice heat engine in the large output power compared to the other similar geometry due to the box shaped transmission function which comes from the electron energy miniband. Our theory predicts that the maximum power under optimized conditions can be larger than similar resonant tunneling devices, with comparable efficiency at maximum power.  Second, a different regime of the system parameters makes the engine works as a refrigerator, and we shows the optimized regime of the parameters for the maximum cooling power and coefficient of performance.  In addition, we analyzed the phonon heat current to find the total efficiency by the performance of electrons and phonons together. The reduction of phonon heat current in the superlattice compared to the corresponding bulk material offers higher total efficiency. The trade off between power and efficiency is overcome by the reduction of the phonon heat current which can be achieved either by operating at low temperatures, or by engineering the system to have low phonon conductivity, while keeping high electron conductivity. Easy fabrication of these devices with advantages for both the power and efficiency show this heat engine is a promising device for next-generation thermoelectrics.


\section*{Acknowledgments}
We would like to thank Antonio Badolato for suggesting this line of research and for discussions.  We thank  Jian-Hua Jiang, and Paul Ampadu for discussions, and Bj\"orn Sothmann and Rafa S\'anchez for helpful comments on the manuscript.  This work is dedicated to the memory of Markus B\"uttiker, a mentor, colleague, and friend.

\end{document}